\documentclass[twocolumn,letterpaper,aps,prc,superscriptaddress,showpacs,nofootinbib,floatfix]{revtex4}

\usepackage{graphicx}	
\usepackage{xspace}	

\newcommand{\pt}{\mbox{$p_T$}\xspace}
\newcommand{\kt}{\mbox{$k_T$}\xspace}
\newcommand{\mt}{\mbox{$m_T$}\xspace}
\newcommand{\RAA}{\mbox{$R_{AA}$}\xspace}
\newcommand{\raa}{\mbox{$R_{AA}$}\xspace}
\newcommand{\rda}{\mbox{$R_{dA}$}\xspace}
\newcommand{\rcp}{\mbox{$R_{\rm cp}$}\xspace}

\newcommand{\Npart}{\mbox{${\langle}N_{\rm part}\rangle$}\xspace}
\newcommand{\Ncoll}{\mbox{${\langle}N_{\rm coll}\rangle$}\xspace}

\newcommand{\sqsn}{\mbox{$\sqrt{s_{_{NN}}}$}\xspace}

\begin{document}


\title{System-size dependence of open-heavy-flavor production 
in nucleus-nucleus collisions at $\sqrt{s_{_{NN}}}$=200~GeV }

\newcommand{\abilene}{Abilene Christian University, Abilene, Texas 79699, USA}
\newcommand{\augie}{Department of Physics, Augustana College, Sioux Falls, South Dakota 57197, USA}
\newcommand{\banaras}{Department of Physics, Banaras Hindu University, Varanasi 221005, India}
\newcommand{\baruch}{Baruch College, City University of New York, New York, New York, 10010 USA}
\newcommand{\bnlcoll}{Collider-Accelerator Department, Brookhaven National Laboratory, Upton, New York 11973-5000, USA}
\newcommand{\bnlphys}{Brookhaven National Laboratory, Upton, New York 11973-5000, USA}
\newcommand{\caucr}{University of California - Riverside, Riverside, California 92521, USA}
\newcommand{\charlesczech}{Charles University, Ovocn\'{y} trh 5, Praha 1, 116 36, Prague, Czech Republic}
\newcommand{\ciae}{Science and Technology on Nuclear Data Laboratory, China Institute of Atomic Energy, Beijing 102413, P.~R.~China}
\newcommand{\cns}{Center for Nuclear Study, Graduate School of Science, University of Tokyo, 7-3-1 Hongo, Bunkyo, Tokyo 113-0033, Japan}
\newcommand{\colorado}{University of Colorado, Boulder, Colorado 80309, USA}
\newcommand{\columbia}{Columbia University, New York, New York 10027 and Nevis Laboratories, Irvington, New York 10533, USA}
\newcommand{\czechtech}{Czech Technical University, Zikova 4, 166 36 Prague 6, Czech Republic}
\newcommand{\dapnia}{Dapnia, CEA Saclay, F-91191, Gif-sur-Yvette, France}
\newcommand{\debrecen}{Debrecen University, H-4010 Debrecen, Egyetem t{\'e}r 1, Hungary}
\newcommand{\elte}{ELTE, E{\"o}tv{\"o}s Lor{\'a}nd University, H - 1117 Budapest, P{\'a}zm{\'a}ny P. s. 1/A, Hungary}
\newcommand{\fit}{Florida Institute of Technology, Melbourne, Florida 32901, USA}
\newcommand{\fsu}{Florida State University, Tallahassee, Florida 32306, USA}
\newcommand{\gsu}{Georgia State University, Atlanta, Georgia 30303, USA}
\newcommand{\hiroshima}{Hiroshima University, Kagamiyama, Higashi-Hiroshima 739-8526, Japan}
\newcommand{\ihepprot}{IHEP Protvino, State Research Center of Russian Federation, Institute for High Energy Physics, Protvino, 142281, Russia}
\newcommand{\illuiuc}{University of Illinois at Urbana-Champaign, Urbana, Illinois 61801, USA}
\newcommand{\inrras}{Institute for Nuclear Research of the Russian Academy of Sciences, prospekt 60\%--letiya Oktyabrya 7a, Moscow 117312, Russia}
\newcommand{\instpasczech}{Institute of Physics, Academy of Sciences of the Czech Republic, Na Slovance 2, 182 21 Prague 8, Czech Republic}
\newcommand{\isu}{Iowa State University, Ames, Iowa 50011, USA}
\newcommand{\jaea}{Advanced Science Research Center, Japan Atomic Energy Agency, 2-4 Shirakata Shirane, Tokai-mura, Naka-gun, Ibaraki-ken 319-1195, Japan}
\newcommand{\jinrdubna}{Joint Institute for Nuclear Research, 141980 Dubna, Moscow Region, Russia}
\newcommand{\kek}{KEK, High Energy Accelerator Research Organization, Tsukuba, Ibaraki 305-0801, Japan}
\newcommand{\korea}{Korea University, Seoul, 136-701, Korea}
\newcommand{\kurchatov}{Russian Research Center ``Kurchatov Institute", Moscow, 123098 Russia}
\newcommand{\kyoto}{Kyoto University, Kyoto 606-8502, Japan}
\newcommand{\labllr}{Laboratoire Leprince-Ringuet, Ecole Polytechnique, CNRS-IN2P3, Route de Saclay, F-91128, Palaiseau, France}
\newcommand{\lahorelums}{Physics Department, Lahore University of Management Sciences, Lahore, Pakistan}
\newcommand{\lawllnl}{Lawrence Livermore National Laboratory, Livermore, California 94550, USA}
\newcommand{\losalamos}{Los Alamos National Laboratory, Los Alamos, New Mexico 87545, USA}
\newcommand{\lpc}{LPC, Universit{\'e} Blaise Pascal, CNRS-IN2P3, Clermont-Fd, 63177 Aubiere Cedex, France}
\newcommand{\lund}{Department of Physics, Lund University, Box 118, SE-221 00 Lund, Sweden}
\newcommand{\michigan}{Department of Physics, University of Michigan, Ann Arbor, Michigan 48109-1040, USA}
\newcommand{\muenster}{Institut f\"ur Kernphysik, University of Muenster, D-48149 Muenster, Germany}
\newcommand{\myongji}{Myongji University, Yongin, Kyonggido 449-728, Korea}
\newcommand{\nagasaki}{Nagasaki Institute of Applied Science, Nagasaki-shi, Nagasaki 851-0193, Japan}
\newcommand{\newmex}{University of New Mexico, Albuquerque, New Mexico 87131, USA }
\newcommand{\nmsu}{New Mexico State University, Las Cruces, New Mexico 88003, USA}
\newcommand{\ohio}{Department of Physics and Astronomy, Ohio University, Athens, Ohio 45701, USA}
\newcommand{\ornl}{Oak Ridge National Laboratory, Oak Ridge, Tennessee 37831, USA}
\newcommand{\orsay}{IPN-Orsay, Universite Paris Sud, CNRS-IN2P3, BP1, F-91406, Orsay, France}
\newcommand{\peking}{Peking University, Beijing 100871, P.~R.~China}
\newcommand{\pnpi}{PNPI, Petersburg Nuclear Physics Institute, Gatchina, Leningrad region, 188300, Russia}
\newcommand{\riken}{RIKEN Nishina Center for Accelerator-Based Science, Wako, Saitama 351-0198, Japan}
\newcommand{\rikjrbrc}{RIKEN BNL Research Center, Brookhaven National Laboratory, Upton, New York 11973-5000, USA}
\newcommand{\rikkyo}{Physics Department, Rikkyo University, 3-34-1 Nishi-Ikebukuro, Toshima, Tokyo 171-8501, Japan}
\newcommand{\saispbstu}{Saint Petersburg State Polytechnic University, St. Petersburg, 195251 Russia}
\newcommand{\saopaulo}{Universidade de S{\~a}o Paulo, Instituto de F\'{\i}sica, Caixa Postal 66318, S{\~a}o Paulo CEP05315-970, Brazil}
\newcommand{\seoulnat}{Seoul National University, Seoul, Korea}
\newcommand{\stonybrkc}{Chemistry Department, Stony Brook University, SUNY, Stony Brook, New York 11794-3400, USA}
\newcommand{\stonycrkp}{Department of Physics and Astronomy, Stony Brook University, SUNY, Stony Brook, New York 11794-3400, USA}
\newcommand{\subatech}{SUBATECH (Ecole des Mines de Nantes, CNRS-IN2P3, Universit{\'e} de Nantes) BP 20722 - 44307, Nantes, France}
\newcommand{\tenn}{University of Tennessee, Knoxville, Tennessee 37996, USA}
\newcommand{\titech}{Department of Physics, Tokyo Institute of Technology, Oh-okayama, Meguro, Tokyo 152-8551, Japan}
\newcommand{\tsukuba}{Institute of Physics, University of Tsukuba, Tsukuba, Ibaraki 305, Japan}
\newcommand{\vandy}{Vanderbilt University, Nashville, Tennessee 37235, USA}
\newcommand{\waseda}{Waseda University, Advanced Research Institute for Science and Engineering, 17 Kikui-cho, Shinjuku-ku, Tokyo 162-0044, Japan}
\newcommand{\weizmann}{Weizmann Institute, Rehovot 76100, Israel}
\newcommand{\wigner}{Institute for Particle and Nuclear Physics, Wigner Research Centre for Physics, Hungarian Academy of Sciences (Wigner RCP, RMKI) H-1525 Budapest 114, POBox 49, Budapest, Hungary}
\newcommand{\yonsei}{Yonsei University, IPAP, Seoul 120\%--749, Korea}
\affiliation{\abilene}
\affiliation{\augie}
\affiliation{\banaras}
\affiliation{\baruch}
\affiliation{\bnlcoll}
\affiliation{\bnlphys}
\affiliation{\caucr}
\affiliation{\charlesczech}
\affiliation{\ciae}
\affiliation{\cns}
\affiliation{\colorado}
\affiliation{\columbia}
\affiliation{\czechtech}
\affiliation{\dapnia}
\affiliation{\debrecen}
\affiliation{\elte}
\affiliation{\fit}
\affiliation{\fsu}
\affiliation{\gsu}
\affiliation{\hiroshima}
\affiliation{\ihepprot}
\affiliation{\illuiuc}
\affiliation{\inrras}
\affiliation{\instpasczech}
\affiliation{\isu}
\affiliation{\jaea}
\affiliation{\jinrdubna}
\affiliation{\kek}
\affiliation{\korea}
\affiliation{\kurchatov}
\affiliation{\kyoto}
\affiliation{\labllr}
\affiliation{\lahorelums}
\affiliation{\lawllnl}
\affiliation{\losalamos}
\affiliation{\lpc}
\affiliation{\lund}
\affiliation{\michigan}
\affiliation{\muenster}
\affiliation{\myongji}
\affiliation{\nagasaki}
\affiliation{\newmex}
\affiliation{\nmsu}
\affiliation{\ohio}
\affiliation{\ornl}
\affiliation{\orsay}
\affiliation{\peking}
\affiliation{\pnpi}
\affiliation{\riken}
\affiliation{\rikjrbrc}
\affiliation{\rikkyo}
\affiliation{\saispbstu}
\affiliation{\saopaulo}
\affiliation{\seoulnat}
\affiliation{\stonybrkc}
\affiliation{\stonycrkp}
\affiliation{\subatech}
\affiliation{\tenn}
\affiliation{\titech}
\affiliation{\tsukuba}
\affiliation{\vandy}
\affiliation{\waseda}
\affiliation{\weizmann}
\affiliation{\wigner}
\affiliation{\yonsei}
\author{A.~Adare} \affiliation{\colorado}
\author{S.~Afanasiev} \affiliation{\jinrdubna}
\author{C.~Aidala} \affiliation{\columbia} \affiliation{\michigan}
\author{N.N.~Ajitanand} \affiliation{\stonybrkc}
\author{Y.~Akiba} \affiliation{\riken} \affiliation{\rikjrbrc}
\author{H.~Al-Bataineh} \affiliation{\nmsu}
\author{J.~Alexander} \affiliation{\stonybrkc}
\author{K.~Aoki} \affiliation{\kyoto} \affiliation{\riken}
\author{N.~Apadula} \affiliation{\stonycrkp}
\author{L.~Aphecetche} \affiliation{\subatech}
% ...
\collaboration{PHENIX Collaboration} \noaffiliation

\date{\today}


\begin{abstract}

The PHENIX Collaboration at the Relativistic Heavy Ion Collider has 
measured open heavy flavor production in Cu$+$Cu collisions at 
$\sqrt{s_{_{NN}}}$=200~GeV through the measurement of electrons at 
midrapidity that originate from semileptonic decays of charm and bottom 
hadrons. In peripheral Cu$+$Cu collisions an enhanced production of 
electrons is observed relative to $p$$+$$p$ collisions scaled by the 
number of binary collisions. In the transverse momentum range from 1 to 
5~GeV/$c$ the nuclear modification factor is $R_{AA}$$\sim$1.4.  As the 
system size increases to more central Cu$+$Cu collisions, the enhancement 
gradually disappears and turns into a suppression. For $p_T>3$~GeV/$c$, 
the suppression reaches $R_{AA}$$\sim$0.8 in the most central collisions. 
The $p_T$ and centrality dependence of $R_{AA}$ in Cu$+$Cu collisions 
agree quantitatively with $R_{AA}$ in $d+$Au and Au$+$Au collisions, if 
compared at similar number of participating nucleons $\langle N_{\rm part} 
\rangle$.

\end{abstract}

	
\maketitle

\section{Introduction}

Studies of the hot matter formed in ultrarelativistic heavy ion collisions, 
such as those produced at the Relativistic Heavy Ion Collider (RHIC), 
require experimental probes, like heavy quarks (charm and bottom), that are 
produced during the collisions.  The temperature of the matter produced in 
$A$$+$$A$ collisions~\cite{Karsch:2001cy,PPG086} is much less than the 
heavy quark masses ($m_{c}$$\sim$1.3~GeV, m$_{b}$ $\sim$ 4.2\,GeV).  Thus 
charm and bottom quarks will only be produced in the initial stage of the 
collision, rather than through thermal excitation or other mechanisms. Once 
produced, they propagate through the hot matter and their kinematic 
distributions are modified through interactions along their path. These 
medium transport effects can be studied experimentally through the spectra 
of decay products from hadrons with heavy quark content.

Indeed, a large suppression of high transverse momentum ($p_T$) electrons 
from semi-leptonic heavy flavor decays was discovered in Au$+$Au collisions at 
$\sqsn = $ 200\,GeV relative to $p$$+$$p$ 
collisions~\cite{PPG077,PPG066,PPG065} in contrast to the predicted reduced 
suppression for heavy quarks related to light quarks due to the ''dead-cone'' 
effect~\cite{dead_cone}.  It is most pronounced in central collisions, i.e. 
collisions with small impact parameters, and disappears in more peripheral 
events. The suppression of heavy flavor at high $p_T$ was widely interpreted 
as evidence that heavy flavor quarks lose a significant amount of energy as 
they traverse hot matter~\cite{Teaney,Gossiaux1,PhysRevC.73.034913}. Many 
effects have been taken into account in various theoretical 
calculations~\cite{Gossiaux2,Gossiaux3,PhysRevC.72.014905,wicks_pqcd,Adil2007139,PhysRevC.74.024902} 
and the data presented here will provide significant additional constraints.

There are many processes that occur in nuclear collisions which can alter 
the kinematic distributions of observed particles.  Modifications of the 
parton distributions in bound nucleons will affect the production rates of 
particles~\cite{EPS09}.  Initial-state parton scattering and energy loss in 
the nucleus will also affect the observed particle 
spectra~\cite{CNM_Eloss}.  

Until recently, the study of $p$$+$$A$ and 
$d$$+$$A$ collisions was considered the best way to investigate and 
quantify these nuclear effects, known generally as cold-nuclear-matter 
effects.  This assumption is now challenged by recent results obtained in 
$p$$+$Pb and $d$$+$Au collisions pointing to additional phenomena that may 
be of hydrodynamic 
origin~\cite{Adare:2013piz,Chatrchyan2013795,Abelev201329,PhysRevLett.110.182302}. 
Therefore, careful comparison of results from $p$$+$$p$, $p$$+$$A$, 
$d$$+$$A$, and $A$$+$$A$, collisions is likely needed to isolate an 
unambiguous signature of hot nuclear matter.

Evidence for such nuclear effects was shown by PHENIX in data from $d$+Au 
collisions at $\sqsn = $ 200\,GeV, where an enhancement of electrons from 
heavy flavor decays was observed relative to $p$$+$$p$ 
collisions~\cite{PPG131} between $1<\pt<5$\,GeV.   The enhancement depends 
on centrality; as the collisions become more central the enhancement becomes 
more and more pronounced, in contrast to the increasing suppression observed 
in Au$+$Au collisions at the same $\sqsn$.  Given these results, it is 
interesting to measure how this enhancement changes over to suppression as 
the system becomes larger.  The system size can be varied by changing the 
colliding nuclei.

At RHIC, effects from cold and hot nuclear matter compete and their 
relative importance likely depends on the system size, which can be 
quantified through the average number of binary nucleon-nucleon collsions 
(\Ncoll) or the average number of participants (\Npart).  Using \Ncoll, 
central $d+$Au collisions show the largest enhancement at $N_{\rm 
coll}$$\sim$15, while central Au$+$Au collisions exhibit the largest 
suppression at $N_{\rm coll}$$\sim$1000 (see Table~\ref{tab:ncoll}).  To 
further investigate system-size dependence, PHENIX has studied Cu$+$Cu 
collisions, also at $\sqsn=$200\,GeV. The \Ncoll range for this 
intermediate-sized system overlaps with $d+$Au as well as Au$+$Au 
collisions and thus Cu$+$Cu allows access to the transition region between 
dominance of enhancement effects and hot-nuclear-matter suppression.

\begin{table}[tbh]
 \caption{\label{tab:NcollNpart}
Values of the average number of binary collisions (\Ncoll) and 
participating nucleons (\Npart) for $d+$Au, Cu$+$Cu, and Au$+$Au 
collisions at $\sqsn = $ 200\,GeV.  \Npart and \Ncoll 
increase with decreasing impact parameter (more central events).
}
 \begin{ruledtabular}
 \begin{tabular}{cccc}
  Colliding species & Centrality & \Ncoll & \Npart     \\
\hline
   $d$+Au  &  0\%--100\%   &  7.59$\pm$0.43 & 9.1$\pm$0.4  \\
                   &  0\%--20\%      &  15.06$\pm$1.01 & 15.4$\pm$1.0   \\
                    &  20\%--40\%    &  10.25$\pm$0.70 & 10.6$\pm$0.7  \\
                   &  40\%--60\%   &  6.58$\pm$0.44 & 7.0$\pm$0.6  \\
                   &  60\%--88\%   &  3.20$\pm$0.19 & 3.1$\pm$0.3  \\
                  \\
   Cu$+$Cu   &  0\%--94\%    &  51.8$\pm$5.6 & 34.6$\pm$1.2 \\
                 &  0\%--10\%    &  182.7$\pm$20.7 & 98.2$\pm$2.4  \\
                 &  0\%--20\%  &  151.8$\pm$17.1 & 85.9$\pm$2.3 \\
                 &  20\%--40\%  &  61.2$\pm$6.6 & 45.2$\pm$1.7 \\
                 &  40\%--60\%  &  22.3$\pm$2.9 & 21.2$\pm$1.4 \\
                 &  60\%--94\%  &  5.1$\pm$0.7 & 6.4$\pm$0.4 \\
   		\\
   Au$+$Au  &  0\%--92\%    &  257.8$\pm$25.4 & 109.1$\pm$4.1 \\
                 &  0\%--10\%    &  955.4$\pm$93.6 & 352.2$\pm$3.3  \\
                 &  10\%--20\%  &  602.6$\pm$59.3 & 234.6$\pm$4.7 \\
                 &  20\%--40\%  &  296.8$\pm$31.1 & 140.4$\pm$4.9 \\
                 &  40\%--60\%  &  90.70$\pm$11.8 & 59.95$\pm$3.6 \\
                 &  60\%--92\%  &  14.50$\pm$4.00 & 14.5$\pm$2.5 \\
 \end{tabular}
 \end{ruledtabular}
 \label{tab:ncoll}
 \end{table}

In this paper we present data of single electrons (we refer to electrons to 
mean the sum of electrons and positrons divided by two) from semi-leptonic 
decays of heavy flavor hadrons obtained in Cu$+$Cu collisions at 
$\sqsn=200$\,GeV. The paper is organized as follows.  
Section~\ref{sec:experiment} presents the experimental setup and describes 
how we measure the inclusive yield of electrons and positrons.  In 
Section~\ref{sec:yield} we discuss how we extract the heavy flavor 
contribution from the inclusive yield.  The results are presented in 
Section~\ref{sec:results} and discussed and compared to previous results in 
Section~\ref{sec:size} and theoretical models in 
Section~\ref{sec:discussion}. Section~\ref{sec:conclusions} gives a summary 
of this work.


\section{Experimental Methods}
\label{sec:experiment}

\begin{figure}[thb]
\includegraphics[width=0.992\linewidth]{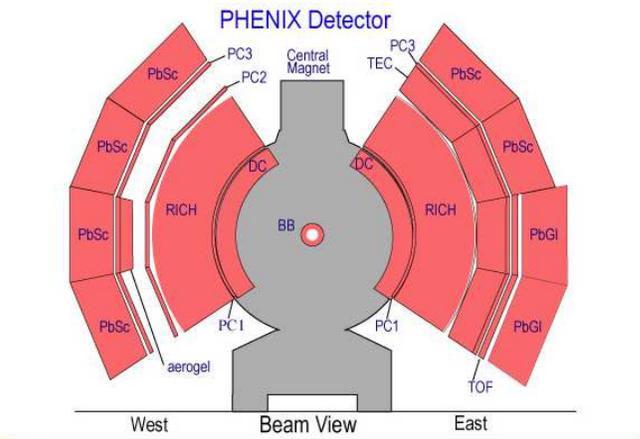}
\caption{(Color online) 
Beam view of the PHENIX central arm detector in the 2005 configuration.
}
\label{fig:detector}
\end{figure}

Figure~\ref{fig:detector} shows the layout of the 2005 PHENIX detector.  
Electrons and positrons are measured using the two central spectrometer 
arms, each of which covers the pseudorapidity range of $|\eta|<0.35$ with 
an azimuthal coverage of $\Delta\phi = \pi/2$.  Charged tracks are 
reconstructed radially outside of an axial magnetic field using layers of 
drift chambers and pad chambers.  Electrons are identified by hits in the 
ring imaging \v{C}erenkov counter (RICH) and by requiring a match between 
an energy deposit in the electromagnetic calorimeter (EMCal) and the 
track's momentum.  The RICH uses CO$_{2}$ gas at atmospheric pressure as 
the \v{C}erenkov radiator.  Electrons and pions begin to radiate in the 
RICH at $\pt>20$\,MeV/c and $\pt>4.9$\,GeV/$c$, respectively.  The EMCal 
comprises four sectors in each arm.  The two lowest sectors of the east arm 
are lead-glass and the remaining six are lead-scintillator.  The angular 
segmentation of the lead-scintillator (lead-glass) is 
$\Delta\phi$x$\Delta\eta$$\sim$0.01x0.01 (0.008x0.008) and the energy 
resolution 
                                       is $\delta$$E/E$$\sim$4.5\%$\oplus$ 
8.3/$\sqrt{E(GeV)}\%$(4.3\% $\oplus$ 
7.7/$\sqrt{E(GeV)}\%$).  A bag filled with He gas at atmospheric pressure 
is placed between the beam pipe and drift chamber (DC) entrance window to 
minimize photon conversions.  Detailed descriptions of the PHENIX detector 
subsystems can be found in~\cite{PHENIXNIM}.

For this analysis two different event samples were used. The first sample 
was obtained with a minimum-bias (MB) trigger, which registers all Cu$+$Cu 
collisions with a coincidence of at least one particle detected in each of 
the two beam-beam counters (BBC). The BBCs are located at $\pm$1.44\,m 
($3.0<|\eta|<3.9$) and comprise 64 \v{C}erenkov counter modules.  Only 
events with a vertex position within$\pm$20\,cm of the nominal z=0 
collision point are kept, giving a sample of 5.08 x 10$^{8}$ events.  A 
second sample was collected with an additional trigger condition requiring 
the detection of an electron candidate in the event. This electron trigger 
(ERT) requires a coincidence between the EMCal and RICH detectors and 
provides an additional 3.3 x 10$^{9}$ events. For the ERT trigger, a 
threshold on the EMCal energy was set at approximately 1.2\,GeV. In our 
analysis we only use electron candidates from this sample above 
$p_T>3$\,GeV/$c$, well beyond the point where the trigger reaches its 
maximium efficiency.  The ERT trigger efficiency for electrons over all 
EMCal sectors was determined to be 67\%$\pm$3\%.  The largest source of 
inefficiency comes from dead trigger tiles.

Centrality is determined by Monte Carlo calculation of the Glauber 
Model~\cite{glauber, Glauber:1970jm} using the measured charge deposited in 
the BBC.  The MB collisions correspond to 0\%--94\% of the inelastic 
cross section.  It is divided into centrality classes covering 0\%--10\%, 
0\%--20\%, 20\%--40\%, 40\%--60\%, and 60\%--94\% of the centrality range (see 
Table~\ref{tab:ncoll}).

The analysis method used here, with some differences, is described in 
detail in~\cite{PPG077}.  Electron candidates start with charged tracks 
reconstructed by the drift chambers and pad chambers.  These tracks are 
then identified as electrons by passing a set of electron identification 
cuts.  First the track is projected to the RICH and at least 5 PMTs 
containing one registered signal are required in a disc (r=11\,cm), with 
an angular size of 0.044\,rad, centered at the projection point.  This 
analysis uses a disc to reduce sensitivity to any possible mirror 
misalignment.  The use of a tight RICH cut ensures a negligible 
contamination of hadrons with \pt above the RICH radiator threshold 
through the \pt range ($<$7\,GeV/$c$) of this analysis.  The track is then 
projected to the EMCal and a three sigma cut is made on the difference 
between the projection and the center of the energy deposition.

A cut is also made on the shape of the EMCal shower, called prob, calculated 
from the deviation between the actual tower energy distribution and the 
expected distribution for an electromagnetic shower and normalized to be 
between 0 and 1.  We require prob$>$0.01 which has a 99\% efficiency for an 
electromagnetic shower while rejecting a large fraction of hadrons.  Finally 
a cut is made on the ratio of the energy deposited in the EMCal to the 
momentum determined by the DC, represented by E/p.  An electron deposits most 
of its energy in the EMCal and because its mass is so small, E$\approx$p and 
E/p for an electron will be close to 1.  The E/p cut is made symmetrically 
around 1 (between 0.8 and 1.2).

Though the electron ID cuts give a good sample of electrons, in a high 
multiplicity environment overlap in the detectors can cause hadrons to be 
misidentified as electrons.  The number and properties of those fake tracks 
reconstructed by random association can be obtained by exchanging, in 
software, the North and South halves of the RICH.  For example, DC tracks 
from the South are matched with the RICH North and vice versa.  After the 
swap, there cannot be any actual tracks, and all reconstructed ones are, by 
definition, fake tracks.  The active area of the North and South RICH 
detectors are identical within $\approx$1\%.  In peripheral collisions, 
3\% of all tracks are mismatches, and in the central collisions that 
fraction rises to 22\%.

A {\sc geant} simulation of the full PHENIX detector was used to determine 
the extrapolation to full azimuthal acceptance and correction for electron 
detection efficiency.  The same eID and fiducial cuts are made on the 
simulation output and the data.  The simulated electrons were generated flat 
in \pt to give sufficient statistics at high momentum, and then weighted 
with a realistic \pt distribution to account for momentum smearing effects 
due to the finite momentum resolution of the drift chamber.


\section{Isolating the Heavy Flavor Yield}
\label{sec:yield}

The inclusive single electron spectrum has contributions from a multitude 
of sources of which heavy flavor decays is only one.  Most of the electrons 
come from decays of light mesons (dominated by the neutral pion Dalitz 
decay, $\pi^{0}\rightarrow \gamma e^{+} e^{-}$)~\cite{PPG084}.  Electrons 
from conversions of decay photons are also significant.  However, the low 
material design of the PHENIX detector minimizes this contribution to less 
than half of that from Dalitz decays.  Direct photons can also be a 
significant contribution to the inclusive electron spectrum, either through 
conversions of real photons in material or manifestations of virtual 
photons as an $e^{+} e^{-}$ pair.  This group of electrons is collectively 
known as ``photonic'' electrons, due to their origins with either a real or 
virtual direct or decay photon.

The other class of electrons, known as ``nonphotonic'', is dominated by the 
decays of open heavy flavor hadrons.  The dielectron decays of the $\rho$, 
$\omega$, and $\phi$ mesons contribute to the inclusive electron sample at 
the few percent level.  Decays of quarkonia, dominated by $J/\psi \rightarrow 
e^{+} e^{-}$~\cite{PPG071}, are a significant source of nonphotonic 
electrons at moderate \pt. Misreconstructed electron tracks from kaon 
$K_{e3}$ 
decays away from the collision vertex are $\sim$10\% of the inclusive 
electrons at $\pt < 1$\,GeV/$c$, but are negligible at higher \pt.  
Electron pairs produced via the Drell-Yan process contribute a negligibly 
small background to the heavy flavor signal.  To isolate the 
contribution of open heavy flavor decays to the inclusive electron spectrum, 
these backgrounds must be determined and removed from the inclusive electron 
sample.  The methods for isolating the open heavy flavor electron yield used 
in this measurement are described in detail in~\cite{PPG077}, and are 
summarized here for completeness.

The first method calculates a cocktail of electrons from the nonheavy flavor 
sources.  Because the PHENIX experiment is a multipurpose detector, most of the 
dominant sources of single electrons have previously been measured in the 
same experiment.  The largest background source comes from the neutral pion, 
both the Dalitz decay and the conversion of photons from the 
$\pi^{0}$$\rightarrow$$\gamma\gamma$ decay.  Using a parametrization of the 
measured $\pi^{0}$ \pt spectra~\cite{PPG084} in a Monte Carlo decay 
generator, the \pt spectrum of daughter electrons is determined.  The \pt 
spectra of the other light mesons that contribute to the cocktail ($\eta$, 
$\rho$, $\omega$, $\eta$', and $\phi$) are derived from the $\pi^{0}$ 
spectrum by $\mt$ scaling (replacing \pt in the parametrization with 
$\sqrt{\pt^{2}+m_{meson}^{2}-m_{\pi^{0}}^{2}}$) and then normalizing to the 
measured meson to pion ratios at high \pt.  At intermediate \pt the 
contribution from J/$\psi$ decays becomes significant and the measured \pt 
spectra~\cite{PPG071} are fit and used as the parent \pt spectra in the 
decay generator.  The cocktail of nonheavy flavor electrons is subtracted 
from the inclusive electron sample to isolate the contribution from open 
heavy flavor electrons.  This method works well at larger \pt where the 
heavy flavor contribution is significant, but suffers from large systematic 
uncertainties at low electron \pt, where the ratio of open heavy flavor 
electrons to all electrons is low.

The second method of isolating the open heavy flavor yield uses a 
``converter'' to deliberately increase the photonic background by a well 
defined amount.  In the standard PHENIX configuration, the number of 
inclusive electrons in a given \pt range $N^{\rm standard}_{e}$ can be 
expressed as

\begin{equation}
N^{\rm standard}_{e} = N^{\gamma} + N^{{\rm non}-\gamma}
\label{eqn:conv_out}
\end{equation}
where $N^{\gamma}$ and $N^{{\rm non}-\gamma}$ are the number of photonic and 
nonphotonic electrons in that \pt bin, respectively.

The converter is a sheet of brass, 0.25\,mm thick, which has a radiation 
length determined to a precision of $\pm$0.25\%.  For a portion of 2005 
running, the converter was wrapped around the beam pipe.  This extra 
material increases the real photonic electron background by an amount 
$R_{\gamma}$, and reduces the nonphotonic electrons by a factor 
$(1-\epsilon)$, giving an inclusive electron yield in the converter 
configuration $N^{\rm converter}_{e}$ of

\begin{equation}
N^{\rm converter}_{e}= R_{\gamma}N^{\gamma} + (1-\epsilon)N^{{\rm non}-\gamma}
\label{eqn:conv_in}
\end{equation}

where the factors $R_{\gamma}$ and $\epsilon$ are determined through 
simulation.  $R_{\gamma}$ has a slight \pt dependence that is prevalent in 
the low \pt region and plateaus at a value of 2.4, and 
$\epsilon=0.021$$\pm$0.005.  The uncertainties on these quantities are 
found by varying the radiation length of the converter material in 
simulation by the uncertainty in the measured converter thickness.

A simultaneous solution of Eqs.~\ref{eqn:conv_out} and~\ref{eqn:conv_in} 
gives the quantity of interest $N^{{\rm non}-\gamma}$.  The remaining 
nonphotonic background electrons are subtracted following the cocktail 
method previously described to isolate the open heavy flavor electron 
contribution.  Because the converter produces an undesirable background for 
other measurements at PHENIX, it is only installed for a relatively short 
amount of time.  Therefore the converter method of background determination 
is limited by the statistics of the data sample taken with the converter 
installed.  However, at low \pt where statistical uncertainties are 
relatively small, the converter method provides meaningful results.  Because 
this region is where the cocktail method is limited by systematic 
uncertainties, and high \pt is where the converter method is limited by 
statistics, the open heavy flavor electron results are determined by the 
converter method at $\pt<1$\,GeV/$c$, and by the cocktail method elsewhere.

\begin{figure}[thb]
\includegraphics[width=1.0\linewidth]{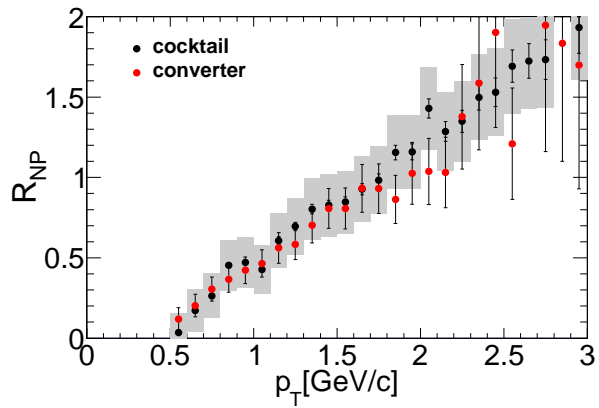}
\caption{(Color online) 
The ratio of nonphotonic to photonic electrons by the converter and 
cocktail methods, for MB Cu$+$Cu collisions.
}
\label{fig:RNP}
\end{figure}

As a stringent cross-check of the methods described here, the ratio of 
nonphotonic electrons to photonic electrons,
\begin{equation}
R_{\rm NP} = \frac{N^{{\rm non}-\gamma}}{N^{\gamma}}
\label{eqn:RNP}
\end{equation}
are compared for the two methods (shown in Fig.~\ref{fig:RNP}).  The gray 
boxes are the systematic uncertainties in determining R$_{\rm NP}$ from the 
cocktail method.

\subsection{Systematic Uncertainties}

The systematic uncertainties on the resulting heavy flavor electron yield 
come from the determination of the inclusive electron yield and the 
uncertainty on the cocktail (converter) method.  The uncertainties are 
explained in detail in~\cite{PPG077} and are summarized here.

The systematic uncertainty on the inclusive yield is a combination of 
three parts: the uncertainty on the run group correction, electron 
identification, and geometric matching.  For the ERT data set there is an 
additional uncertainty that comes from determining the trigger efficiency.  
The run group correction uncertainty comes from the fluctuation of the 
average number of electrons per event (${\langle}N_e/N_{\rm evt}\rangle$) 
for each run (where a run is defined 
as the data taken between successive starts of the PHENIX data acquisition 
system).  The uncertainty on ${\langle}N_e/N_{\rm evt}\rangle$ was 
found to be 1\% and is assigned as the run-group-correction fluctuation.  
The uncertainty on identifying electrons comes from the inability to 
perfectly model the detector in simulation.  It is estimated by repeating 
the acceptance$*$efficiency calculation for tighter and looser cuts which 
are then applied to inclusive yields made with the same cuts.  The ratio 
between the standard and tight or loose cuts is found to be 6\% and is 
taken as the systematic uncertainty.  Mismatching in the detector 
acceptance between simulation and data is an additional uncertainty and 
was found to be 4\%.  The ERT data set is only used in the high \pt region 
where the trigger efficiency is at the plateau value and so the only 
uncertainty is due to the determination of the trigger plateau, 2\%.  The 
total systematic uncertainty on the inclusive MB (ERT) yield is the 
quadrature sum of the previously discussed uncertainties and is found to 
be 7.3 (7.5)\%.

The dominant systematic uncertainty on the cocktail comes from the 
uncertainty on the 2005 Cu$+$Cu neutral pion data~\cite{PPG084} that is used as 
the input parent spectra for all of the light mesons.  The pion data are 
moved up and down by their systematic uncertainties and refit.  These new 
fits are then input into the decay generator and the output decay spectra 
become the upper and lower spread of the systematic uncertainties.  The 
systematic uncertainty on the J/$\psi$ spectra is done in the same manner as 
the pions.  The rest of the light mesons are moved up and down by the 
uncertainty on the meson/$\pi^{0}$ ratios at high \pt.  The systematic 
uncertainty on the conversion yield is found by scaling the conversion 
probability up and down by 10\%.  This gives a conservative estimate of the 
uncertainty on the amount of conversion material within PHENIX.  The 
$K_{e3}$ is 
assigned a 50\% uncertainty as in previous analyses.  The systematic 
uncertainty on the cocktail is dependent on \pt and centrality but has an 
average value of $\sim$12\% for MB collisions.

The systematic uncertainty on the converter analysis comes from two 
sources:  the already described uncertainty on the inclusive yield and the 
uncertainty derived from extracting a nonphotonic yield from the converter 
analysis.  These uncertainties are independent and added in quadrature.  
R$_{\gamma}$, $\epsilon$, and N$_{inc}^{C}$ are moved up and down by their 
systematic uncertainties and the effect on the yield is calculated and 
then added in quadrature.  The overall converter systematic uncertainty is 
found to be 8\% for MB.  Table~\ref{tab:sys} summarizes the different 
systematic uncertainties.

\begin{table}[tbh]
\caption{
Systematic uncertainties on the determination of the open heavy flavor yield 
of electrons for MB collisions.}
 \begin{ruledtabular}
 \begin{tabular}{lc}
  Run Group Correction          & 1\% \\
  Acceptance$\ast$Efficiency    & 6\% \\
  Geometric Matching            & 4\% \\
  Trigger Efficiency            & 2\%\\
	MB (ERT) Inclusive Yield & 7.3\% (7.5\%)\\
	Cocktail (Average) & 12\%\\
	Converter & 8\%\\
 \end{tabular}
 \end{ruledtabular}
 \label{tab:sys}
 \end{table}


\section{Results from Cu$+$Cu Collisions}
\label{sec:results}

The invariant yield of heavy flavor electrons is calculated as a function of 
\pt using the following formula:
\begin{equation}
\frac{1}{2\pi\pt}\frac{d^{2}N^{e}}{d\pt dy} = \frac{1}{2\pi\pt 
N_{events}}\frac{N_{e_{\rm HF}}}{2}\frac{1}{\Delta\pt\Delta y}
\frac{1}{\epsilon_{\rm BBC}\epsilon_{eID}},
\label{eq:yield}
\end{equation}
where N$_{events}$ is the number of events, $\Delta\pt$ is the \pt bin 
width, $\Delta$y is the rapidity range ($|y|<0.35$), $\epsilon_{\rm BBC}$ is 
the BBC efficiency for MB (94\%), $\epsilon_{eID}$ is acceptance 
and efficiency correction, and N$_{e_{\rm HF}}$ is the calculated number of heavy 
flavor electrons and positrons from either the cocktail or converter method.

When plotting the invariant yield vs \pt, the average value is plotted at 
the bin center.  However, for a steeply falling spectrum, the average value 
does not lie at the center of the bin.  This is corrected by adjusting the 
average value over the bin to correspond to the value of the yield at the 
\pt bin center.  This procedure assumes that the invariant yield as a 
function of \pt varies smoothly, which is a reasonable assumption.

\begin{figure}[thb]
\includegraphics[width=0.992\linewidth]{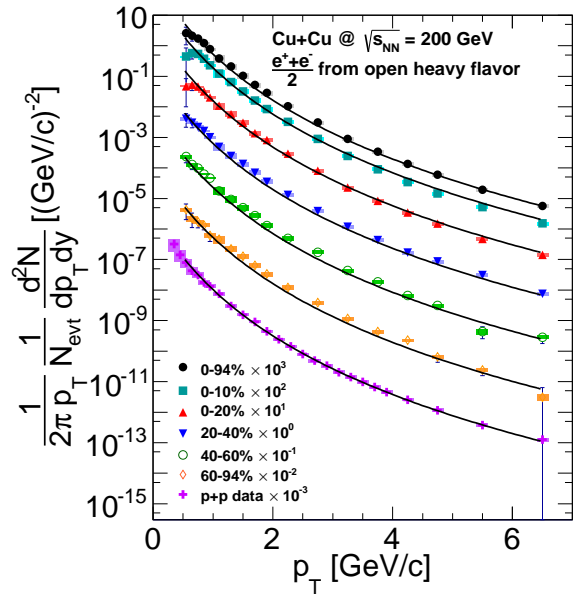}
\caption{(Color online) 
The \pt spectra of electrons from the decays of open heavy flavor hadrons 
produced in Cu$+$Cu collisions, separated by centrality.  The lines are a 
fit to the $p$$+$$p$ data~\cite{PPG077} scaled by $\Ncoll$.
}
\label{fig:CuCu_spectra}
\end{figure}

The \pt spectra of heavy flavor electrons ($e_{\rm HF}$) produced in Cu$+$Cu 
collisions at $\sqsn = $ 200\,GeV are shown for 5 different centralities in 
Fig.~\ref{fig:CuCu_spectra}, along with a fit to the $e_{\rm HF}$ spectrum from 
$p$$+$$p$ collisions (as reported in~\cite{PPG077}) scaled by \Ncoll.  
Above $\pt = 1$\,GeV/$c$, the spectra are taken from the cocktail method 
described previously, while at lower \pt the spectra are determined by the 
converter method.

\begin{figure*}[thb]
\includegraphics[width=0.97\linewidth]{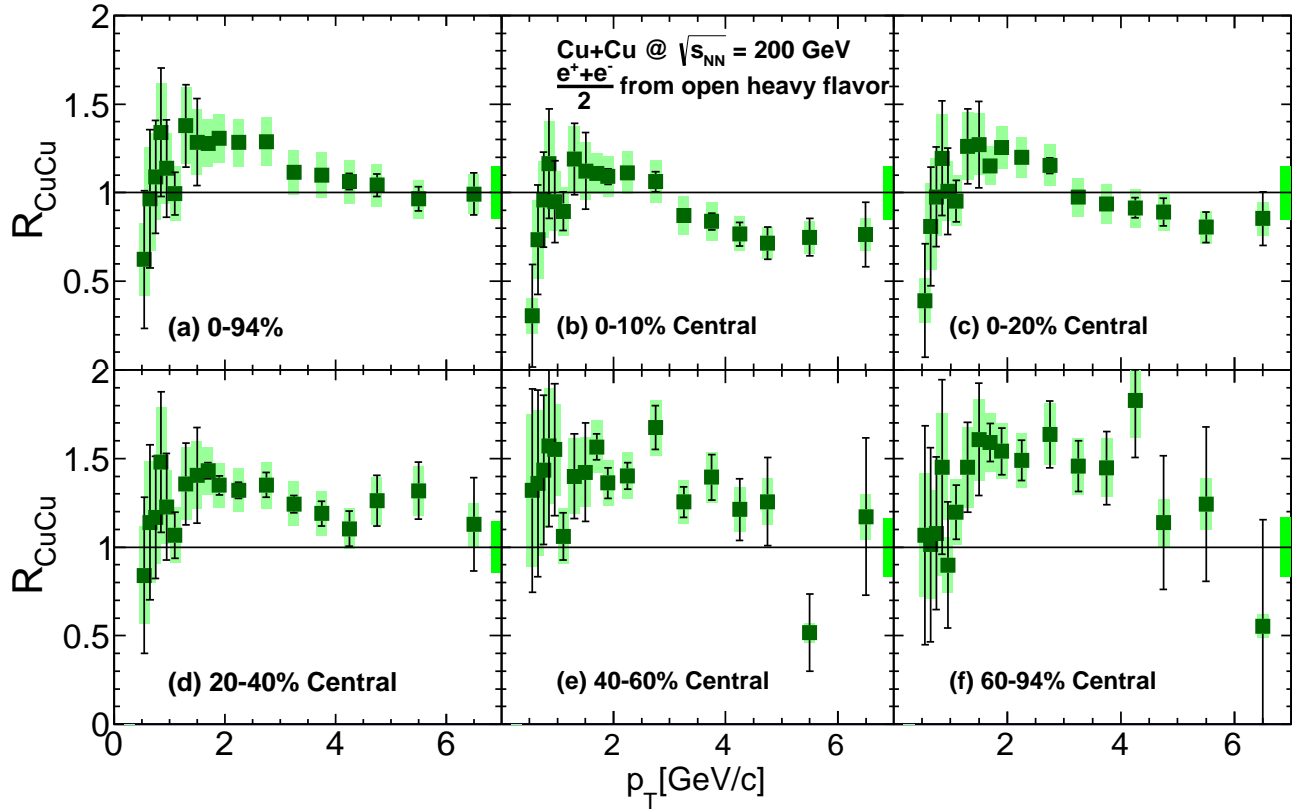}
\caption{(Color online) 
The nuclear modification factor for MB (0\%--94\%) and five centrality 
bins (0\%--10\%, 0\%--20\%, 20\%--40\%, 40\%--60\%, 60\%--94\%).  The 
boxes around one are Type C global uncertainties, which include the \Ncoll
scaling error and $p$$+$$p$ global uncertainty (9.9\%) added in 
quadrature.
}
\label{fig:RCuCu_cp}
\end{figure*}

To quantify nuclear effects, the nuclear modification factor $\raa$ is 
calculated according to: \begin{equation} \raa = 
\frac{dN^{e}_{A+A}/dp_T}{\langle N_{\rm coll} \rangle \times 
dN^{e}_{p+p}/dp_T}, \label{eqn:rda} \end{equation} where $dN^e_{A+A}/dp_T$ 
($dN^{e}_{p+p}/dp_T$) is the differential yield in $A$$+$$A$ ($p$$+$$p$) 
collisions. An $\raa$ value of one indicates that the A+A data are well 
described by a superposition of independent $p$$+$$p$ collisions.  
Following~\cite{PPG077, PPG131}, at $\pt<1.6$\,GeV/$c$, $\raa$ is 
calculated by dividing the Cu$+$Cu spectra by the $p$$+$$p$ spectra 
point-by-point. The statistical (systematic) uncertainties on $\raa$ in 
this range are the quadrature sum of the statistical (systematic) 
uncertainties on the Cu$+$Cu and $p$$+$$p$ yields in a given \pt bin.

Above \pt = 1.6\,GeV/$c$, where the $p$$+$$p$ data 
are well represented by the shape from fixed-order plus 
next-to-leading-log calculations from Ref.~\cite{FONLL}, a fit to 
that shape is used to represent the $p$$+$$p$ denominator.  A function of 
the form \begin{equation} Y(\pt) = \frac{A}{(\pt +B)^{n}} 
\label{eqn:pp_fit} \end{equation} is fit to these data, where 
A=0.0067$\pm$0.0035\,(GeV/$c$)$^{-2}$, B=1.079$\pm$0.085\,GeV/$c$, and 
n=8.86$\pm$0.23.  Here, the statistical uncertainty on $\raa$ is 
determined by the statistical uncertainty on the Cu$+$Cu spectra.  The 
systematic uncertainty on $\raa$ is the quadrature sum of the systematic 
uncertainty on the $e_{\rm HF}$ yield from Cu$+$Cu and $p$$+$$p$, and the 
statistical uncertainty on the fit to the $p$$+$$p$ data.  The Type C 
global scaling uncertainty plotted around 1 is the quadrature sum of the 
global uncertainty on the $p$$+$$p$ spectra and the uncertainty on 
$\Ncoll$.

Figure~\ref{fig:RCuCu_cp}(b) shows the nuclear modification factor for the 
0\%--10\% most central Cu$+$Cu collisions, in which a moderate suppression of 
$e_{\rm HF}$ is observed for $\pt>3$\,GeV/$c$.  This suppression is usually 
attributed to energy loss in the hot nuclear medium.  Although this is a 
significant deviation from a superposition of independent $p$$+$$p$ 
collisions, the magnitude of suppression is smaller than what is seen in 
central Au$+$Au collisions~\cite{PPG077,PPG066}.

\begin{figure}[thb]
\includegraphics[width=1.0\linewidth]{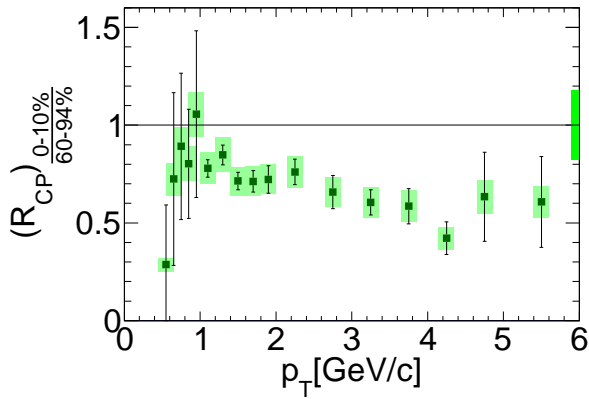}
\caption{(Color online) 
The ratio $\rcp$ of the most central 0\%--10\% $e_{\rm HF}$ spectra to the 
most peripheral 60\%--94\%, scaled by $\Ncoll$.  Type C uncertainty is the 
uncertainty on the determination of $\Ncoll$ for each centrality, shown as 
a box around 1.  
}
\label{fig:Rcp}
\end{figure}

In contrast, a significant enhancement is observed in more peripheral Cu$+$Cu 
collisions, Fig.~\ref{fig:RCuCu_cp}.  To quantitatively examine the 
difference within the Cu$+$Cu system itself, the $\Ncoll$-scaled ratio of the 
most central to most peripheral spectra $\rcp$, defined as
\begin{equation}
\rcp = \frac{N_{\rm coll}^{\rm peripheral}}
{N_{\rm coll}^{\rm central}}  \times  
\frac{dN^{\rm central}_{\rm Cu+Cu}/dp_{T}}{dN^{\rm peripheral}_{\rm Cu+Cu}/dp_{T}}
\label{eqn:rcp}
\end{equation}
and is shown in Fig.~\ref{fig:Rcp}.  Most of the systematic uncertainties 
cancel in $\rcp$, leaving only the uncertainty on the centrality dependent 
cocktail and the ratio of \Ncoll values.  A clear suppression is seen in 
the most central collisions relative to the most peripheral, which can be 
attributed to the suppression effects of the hot, dense partonic matter 
dominating in central collisions.

\begin{figure}[htb]
\includegraphics[width=0.992\linewidth]{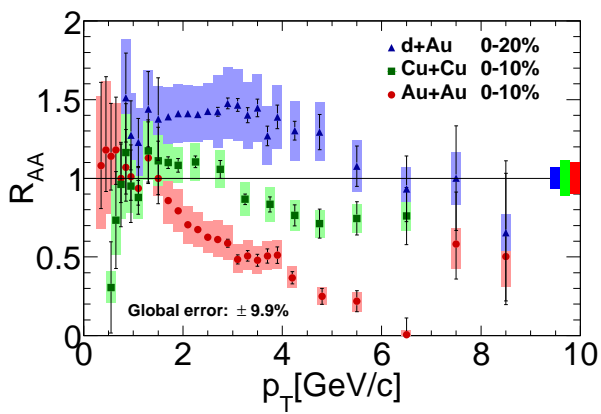}
\caption{(Color online) 
The nuclear modification factors for $e_{\rm HF}$ at midrapidity in 
central $d$+Au~\cite{PPG131}, Cu$+$Cu, and Au$+$Au~\cite{PPG077} 
collisions at $\sqsn = $ 200\,GeV.  The boxes around one are Type C 
uncertainties, which include the $\Ncoll$ scaling error.  The global 
uncertainty is that on the $p$$+$$p$ yield.
}
\label{fig:all}
\end{figure}

\begin{figure}[thb]
\includegraphics[width=0.992\linewidth]{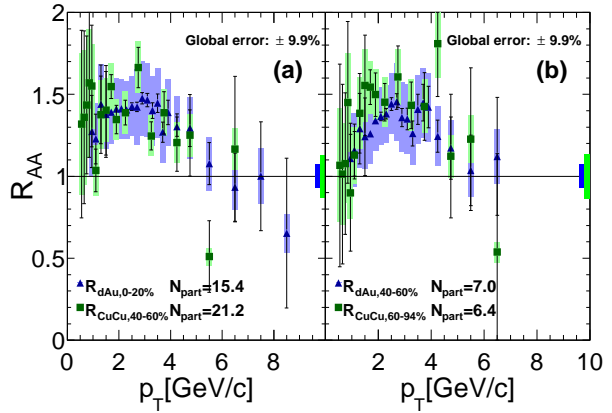}
\caption{(Color online) 
The nuclear modification factors for 0\%--20\% $d+$Au~\cite{PPG131} and 
40\%--60\% Cu$+$Cu collisions.  Right:  The nuclear modification factors 
for 40\%--60\% $d+$Au and 60\%--94\% Cu$+$Cu collisions.  The boxes around 
one are Type C uncertainties, which include the $\Ncoll$ scaling error.  
The global uncertainty is that on the $p$$+$$p$ yield.
}
\label{fig:RCuRdA}
\end{figure}


\begin{figure}[htb]
\includegraphics[width=0.992\linewidth]{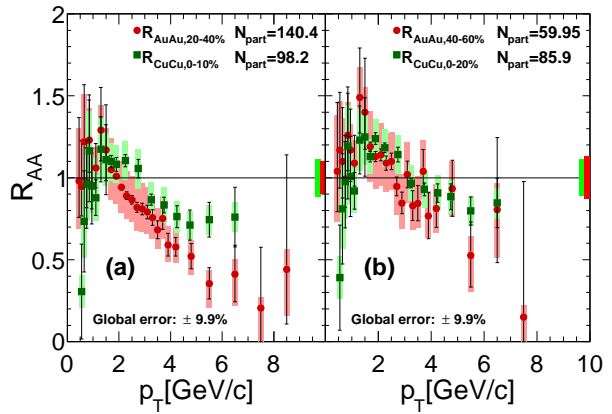}
\caption{(Color online) 
The nuclear modification factors for (a) 0\%--10\% Cu$+$Cu and 20\%--40\% 
Au$+$Au~\cite{PPG077} collisions and (b) 0\%--20\% Cu$+$Cu and 40\%--60\% 
Au$+$Au collisions.  The boxes around one are Type C uncertainties, which 
include the $\Ncoll$ scaling error.  The global uncertainty is the global 
uncertainty on the $p$$+$$p$ yield.
}
\label{fig:RCuRAA}
\end{figure}

\begin{figure*}[thb]
\includegraphics[width=0.995\linewidth]{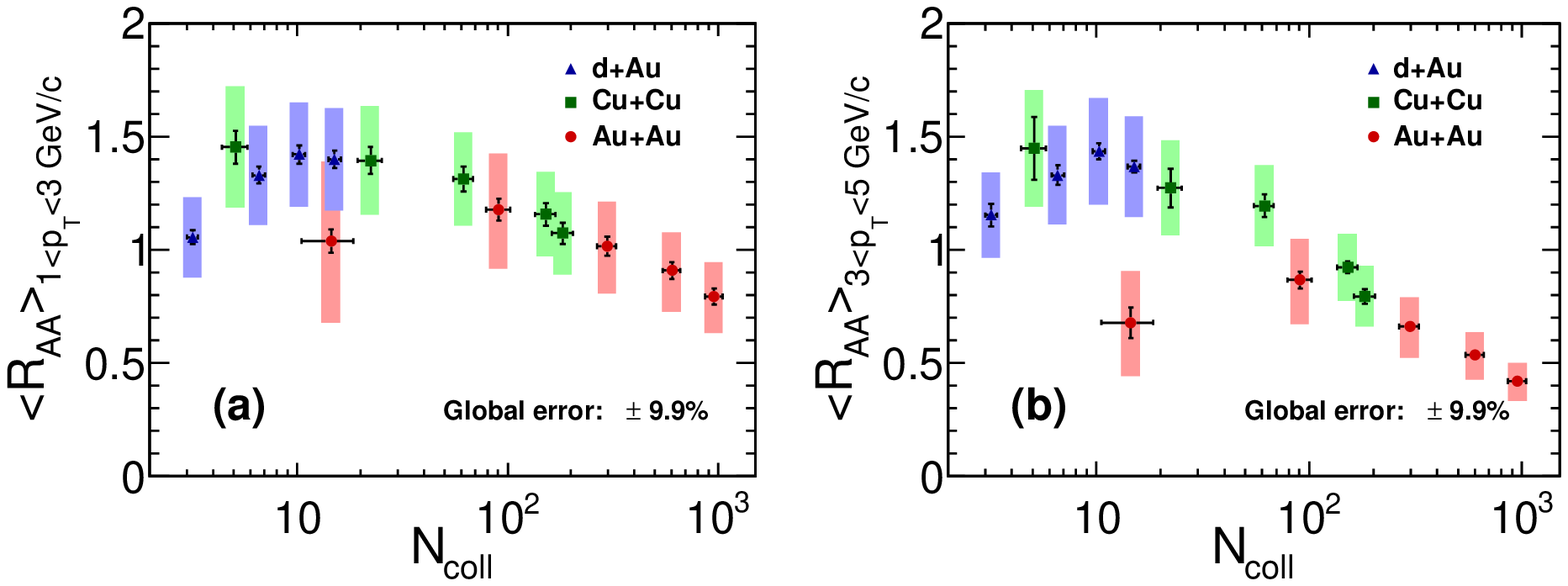}
\caption{(Color online) 
The nuclear modification factors, averaged over $1<\pt<3$\,GeV/$c$ (a) 
and $3<\pt<5$\,GeV/$c$ (b), for $e_{\rm HF}$ at midrapidity in 
$d$+Au~\cite{PPG131}, Cu$+$Cu, and Au$+$Au~\cite{PPG077} collisions plotted 
as a function of $\Ncoll$. 
}
\label{fig:RAA_vs_Ncoll}
\end{figure*}

\begin{figure*}[htb]
\includegraphics[width=0.995\linewidth]{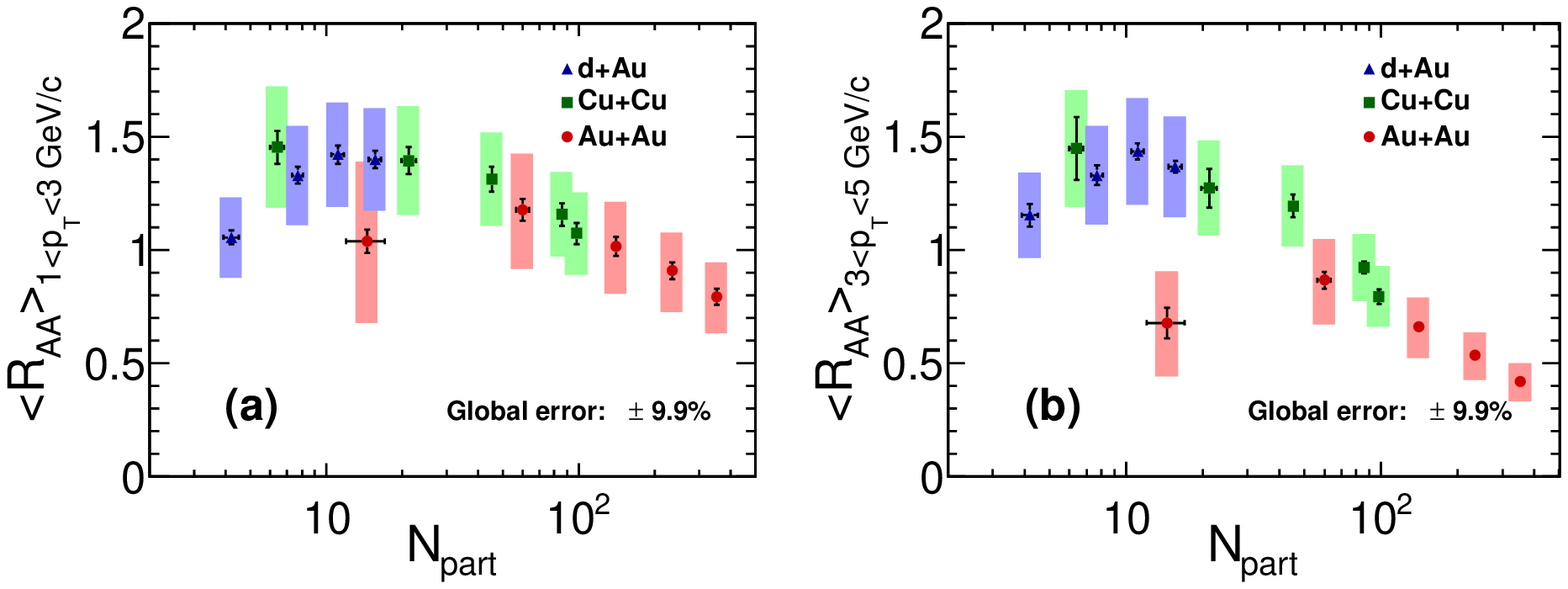}
\caption{(Color online) 
The nuclear modification factors, averaged over $1<\pt<3$\,GeV/$c$ (a) 
and $3<\pt<5$\,GeV/$c$ (b), for $e_{\rm HF}$ at midrapidity in 
$d$+Au~\cite{PPG131}, Cu$+$Cu, and Au$+$Au~\cite{PPG077} collisions plotted 
as a function of $\Npart$.  
}
\label{fig:RAA_vs_Npart}
\end{figure*}

\begin{figure}[thb]
\includegraphics[width=0.992\linewidth]{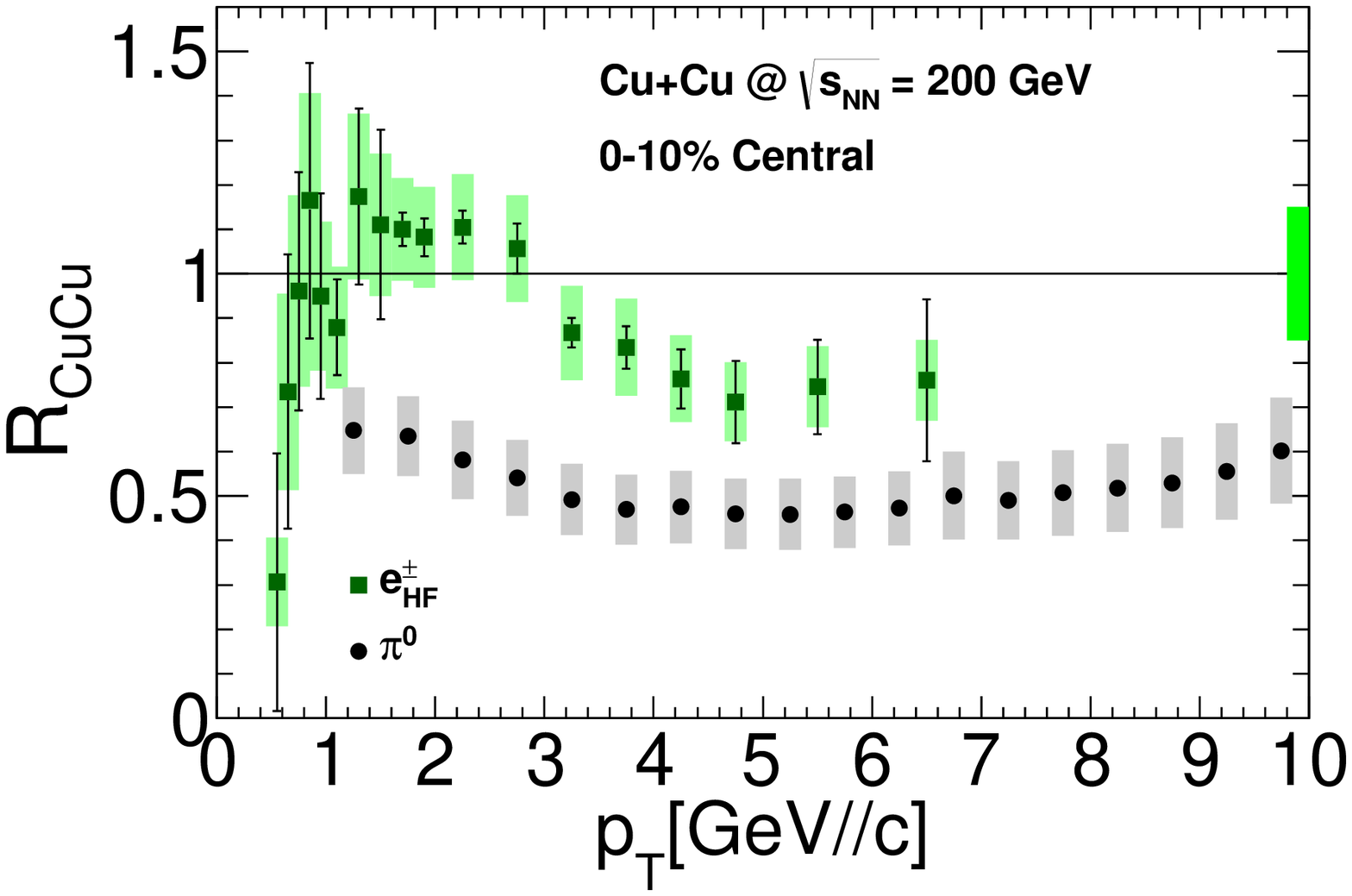}
\caption{(Color online) 
$\raa$ for $\pi^{0}$ and $e_{\rm HF}$. The boxes around one are Type C 
global uncertainties, which include the $\Ncoll$ scaling uncertainty and 
$p$$+$$p$ global uncertainty.
}
\label{fig:RAA_pi0}
\includegraphics[width=0.992\linewidth]{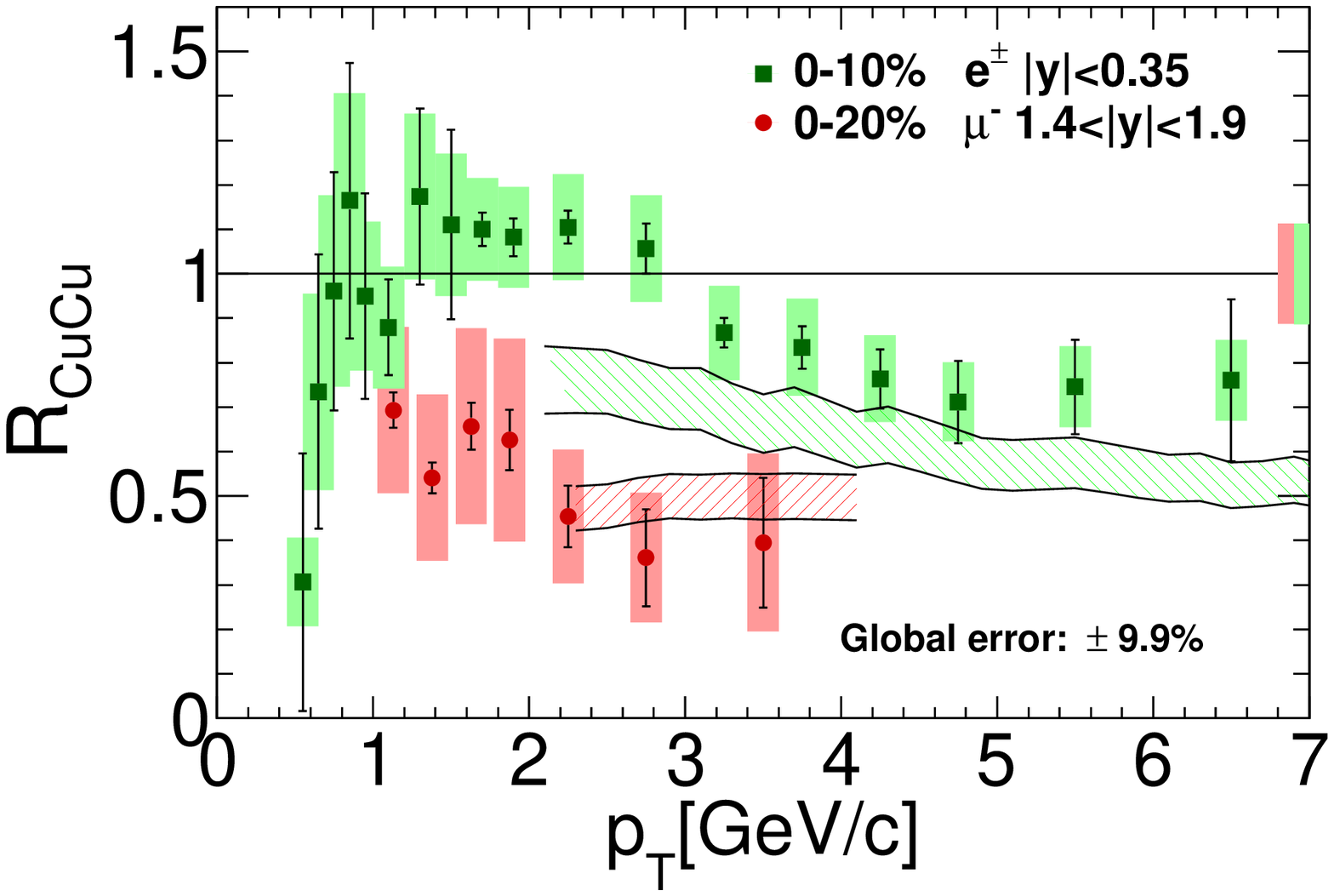}
\caption{(Color online) 
The nuclear modification factors for $e_{\rm HF}$ at midrapidity and 
$\mu^{-}_{\rm HF}$ at forward rapidity, for the 0\%--20\% most central 
Cu$+$Cu collisions.  The boxes around one are Type C uncertainties from 
the \Ncoll scaling error.  The global uncertainty is that on the $p$$+$$p$ 
yield.  Model bands are calculations from~\cite{SharmaVitev} including 
partonic energy loss, energy loss from fragmentation and dissociation, and 
effects from nuclear matter.
}
\label{fig:cucu_raa_vitev}
\end{figure}

\section{System Size Dependence}
\label{sec:size}

The full extent of the system size dependence is directly illustrated by 
comparing the most central bins of all three systems in Fig.~\ref{fig:all}.  
There is a clear enhancement in central $d$+Au collisions, which gives way to 
a slight suppression in central Cu$+$Cu collisions, and finally a large 
suppression in the most central Au$+$Au bin.

If results from different systems are compared in centrality bins of 
comparable system size the trend is similar.  Here we take the number of 
nucleons participating in the collision, \Npart, as a measure of the 
centrality and of the size of the system.  The centrality selections are 
the same if \Ncoll is used as a measure of the system size instead.  
Figure~\ref{fig:RCuRdA} shows overlays of the $\RAA$ for peripheral 
Cu$+$Cu collisions with the $\rda$ for $d+$Au collisions at a comparable 
value of \Npart.  A similar enhancement is seen for the two systems.


Within the Cu$+$Cu system, the enhancement is overtaken by suppression as 
the average impact parameter decreases and with it the number of 
collisions increases.  To compare the levels of suppression in Cu$+$Cu and 
Au$+$Au collisions, the nuclear modification factors for heavy flavor 
electrons in centrality classes with comparable $\Npart$ values are shown 
in Fig.~\ref{fig:RCuRAA}.  Here our centrality selections do not allow for 
as close a match, but a similar level of modification is seen for the 
different systems at similar values of $\Npart$.

Rather than comparing $\RAA$ vs \pt for similar system size, one can also 
compare average $\RAA$ values in a given \pt range as a function of 
$\Npart$ or $\Ncoll$.  The average value of the nuclear modification factor 
for $1<\pt<3$\,GeV/$c$ and $3<\pt<5$\,GeV/$c$ for the three collision 
species 
is shown in Figs.~\ref{fig:RAA_vs_Ncoll} and \ref{fig:RAA_vs_Npart}, as a 
function of $\Ncoll$ and $\Npart$, respectively.  With the exception of the 
most peripheral Au$+$Au bin in the higher \pt range, a trend of increasing 
enhancement followed by suppression is seen among the three distinct systems, 
with Cu$+$Cu showing evidence of both. This common trend suggests that the 
enhancement and suppression effects are dependent on the size of the 
colliding system and the produced medium.

\section{Discussion}
\label{sec:discussion}

The heavy quark data from Cu$+$Cu collisions display enhancement and 
suppression features similar to those found in both $d+$Au and Au$+$Au 
collisions, respectively.  The enhancement seen for heavy flavor electrons 
in central $d+$Au is larger than what is observed for pions and kaons at 
the same collision energy~\cite{PPG044}.  In central $d$+Au collisions, a 
mass-dependent enhancement is observed for identified pions, kaons, 
and protons~\cite{PPG146}.  The proton spectra show the largest 
enhancement and reach an $\rda$ of $\sim$1.5 at \pt$=$3\,GeV/$c$ in MB 
$d$+Au collisions.  This hardening of hadron spectra in nuclear collisions 
compared to $p$$+$$p$ collisions is known as the ``Cronin 
effect''~\cite{Cronin} and is used generically for \raa$>$1 
observations.  

Early explanations of the mechanism behind the Cronin effect 
relied on $\kt$ boosts to partons via scattering in the nucleus before the 
hard scattering and subsequent fragmentation~\cite{Accardi}; however, this 
hypothesis does not explain the observed mass dependence, because the 
$\kt$ transverse momentum kicks in the nucleus presumably occur before 
hadronization and therefore could not preferentially boost protons more 
than pions.  An alternative scenario involving recombination of soft 
partons in the hadronization process naturally gives a difference between 
meson and baryon enhancement~\cite{Hwa1,Hwa2}, but it is not immediately 
clear what effect this has on heavy flavor electrons, which are from a 
mixture of charm and bottom meson and baryon decays (though most are from 
mesons).  The baryon enhancement observed in $d$+Au and $A$$+$$A$ 
collisions at RHIC can also suppress $e_{\rm HF}$ production at moderate 
\pt, because charmed baryons have a smaller branching ratio to electrons 
than charmed mesons~\cite{charm_chemistry}; however, currently no 
measurements of charmed baryons at RHIC energies exist to confirm any 
changes in the charmed hadron chemistry.

Charm and bottom production at midrapidity is dominated by gluon fusion 
and samples nuclear $x$-values of $\sim$10$^{-2}$, where modification of 
the gluon PDF may be significant in central collisions.  Because the 
observed enhancement occurs in peripheral Cu$+$Cu collisions with a large 
average impact parameter, where the spatially-dependent nuclear PDF is 
expected to have minimal changes from the free-nucleon PDF~\cite{Helenius}, 
this may suggest that gluon modification is not the dominant effect at 
midrapidity.  Parton energy loss in the nucleus may also affect heavy 
flavor production in nuclear collisions~\cite{CNM_Eloss}. These effects 
are also expected to occur in the initial stages of central nuclear 
collisions, prior to the formation of the hot nuclear medium.
The observed enhancement is theoretically unexplained.

Several mechanisms have been put forth to explain the large $e_{\rm HF}$ 
suppression in Au$+$Au collisions (shown in Fig.~\ref{fig:RCuRAA}) when it was 
found that radiative energy loss alone was not sufficient to reproduce the 
suppression~\cite{Wicks2007426}.  Recent models involving collisional energy 
loss and energy loss through in-medium dissociation of heavy flavor mesons in 
addition to gluon radiation have proven more successful at describing the 
Au$+$Au 
data~\cite{zhang_pqcd,wicks_pqcd,Adil2007139,Dominguez2009246,Dominguez200999}.  
Fragmentation and dissociation have recently been used to describe the 
suppression of the quarkonia yield in Au$+$Au collisions and could also be 
applied to the heavy-light bound states of the D and B 
meson~\cite{PhysRevC.72.034906}. These effects are sensitive to the formation 
times of the mesons and the hot nuclear medium.

It was originally thought that heavy quarks would exhibit less suppression 
than light quarks in a deconfined medium, due to a suppression of small-angle 
gluon radiation known as the ``dead cone'' effect~\cite{dead_cone}.  The 
$e_{\rm HF}$ and $\pi^{0}$ $\raa$ for central Cu$+$Cu collisions are shown in 
Fig.~\ref{fig:RAA_pi0}.  The heavy flavor electrons seem to approach the 
level of suppression of the neutral pions, though the electron \pt range is 
limited.  While this may suggest a difference in energy loss for light and 
heavy quarks, the peripheral $\pi^{0}$ data show none of the enhancement that 
is present for heavy flavor.  Because the nuclear effects are expected to be 
present in the initial state of central collisions, the different level of 
suppression for $e_{\rm HF}$ and $\pi^{0}$ may indicate that the initial state 
effects on light and heavy quarks are different.  A similar difference is 
also observed in $d+$Au collisions, where the $e_{\rm HF}$ show significant 
enhancement while the $\pi^{0}$ does not.

Previous PHENIX measurements at forward rapidity ($1.4<y<1.9$) showed a 
significant suppression of heavy flavor muons ($\mu_{\rm HF}$) in central 
Cu$+$Cu collisions~\cite{PPG117}.  The magnitude of this suppression at 
forward rapidity in Cu$+$Cu (shown in Fig.~\ref{fig:cucu_raa_vitev}) is 
comparable to the suppression of $e_{\rm HF}$ in central Au$+$Au 
collisions at midrapidity.  This observation is difficult to reconcile 
with explanations of heavy flavor suppression that depend solely on energy 
loss in the hot nuclear medium, because the energy density of the matter 
created in central Au$+$Au collisions is expected to be larger than in 
Cu$+$Cu collisions~\cite{Bjorken, PPG117}.  Because open heavy flavor is 
significantly more suppressed at forward rapidity than at midrapidity in 
Cu$+$Cu, additional nuclear effects, such as gluon shadowing at low $x$ or 
partonic energy loss in the nucleus, may be significant.  Suppression 
through shadowing effects may also be relevant to heavy flavor production 
at midrapidity at the Large Hadron Collider~\cite{ALICE}, because the 
$\sqsn$ is higher than at RHIC and probes a lower $x$-range 
within the nucleus at midrapidity.

The heavy flavor electrons and muons are compared in 
Fig.~\ref{fig:cucu_raa_vitev} to a theoretical prediction that combines the 
effects of partonic energy loss, energy loss from fragmentation and 
dissociation, and includes nuclear matter effects such as shadowing and 
Cronin enhancement due to parton scattering in the 
nucleus~\cite{SharmaVitev}.  While consistent within uncertainties, the model 
predicts more suppression for heavy flavor electrons than seen in the data.  
The B mesons are heavier and so dissociation is the dominant contribution to 
the energy loss for the entire \pt range at RHIC in this model.  On the 
other hand, with its lighter mass, the D meson transitions at 
\pt$\sim$5\,GeV to the traditional partonic energy loss.  However, it is 
critical to test models against the full range of system sizes to have 
confidence in the underlying model physics and so calculations are needed for 
$d$+Au and peripheral Cu$+$Cu.


\section{Summary and Conclusions}
\label{sec:conclusions}

The Cu$+$Cu data presented here build a bridge between the enhancement observed 
in $d+$Au collisions and the suppression found in Au$+$Au. We find that for 
electrons between 1 and 3\,GeV/$c$ the variation in $\raa$ is common as a 
function of \Npart or \Ncoll in $d$+Au, Cu$+$Cu, and Au$+$Au.  For electrons 
between 3--5\,GeV/$c$ this relation also holds with the exception of the most 
peripheral Au$+$Au.  Peripheral collisions of Cu nuclei display an enhancement 
of open heavy flavor at moderate \pt that is consistent with the 
enhancement observed in $d$+Au collisions at similar values of $\Ncoll$ and 
$\Npart$, which suggests significant effects on heavy quark production are 
present in the initial state of heavy ion collisions.  In central Cu$+$Cu 
collisions, open heavy flavor at midrapidity is moderately suppressed when 
compared to a superposition of independent $p$$+$$p$ collisions, and 
significantly suppressed compared to peripheral Cu$+$Cu collisions. The nuclear 
modification factor $\raa$ displays a suppression that is consistent with 
that seen in semi-peripheral Au$+$Au collisions with a similar system size, 
suggesting that the suppressing effects from hot nuclear matter are becoming 
dominant.

While partonic energy loss in medium alone does not describe either the 
Cu$+$Cu or Au$+$Au $e_{\rm HF}$ data, a model which incorporates initial 
state gluon shadowing, parton scattering and energy loss in nuclear matter, 
followed by dissociative energy loss in the hot medium, gives a reasonable 
description of central Cu$+$Cu open heavy flavor data at both midrapidity 
and forward rapidity.  Models that describe central Au$+$Au should also be 
tested against the Cu$+$Cu and $d$+Au data.  A number of different effects 
must be balanced to describe the data, which demonstrates the complicated 
interplay of effects from nuclear matter and those from the hot medium in 
heavy ion collisions.


\section*{ACKNOWLEDGMENTS}


We thank the staff of the Collider-Accelerator and Physics
Departments at Brookhaven National Laboratory and the staff of
the other PHENIX participating institutions for their vital
contributions.  We acknowledge support from the 
Office of Nuclear Physics in the 
Office of Science of the Department of Energy, the
National Science Foundation, 
Abilene Christian University Research Council, 
Research Foundation of SUNY, and 
Dean of the College of Arts and Sciences, Vanderbilt University 
(U.S.A),
Ministry of Education, Culture, Sports, Science, and Technology
and the Japan Society for the Promotion of Science (Japan),
Conselho Nacional de Desenvolvimento Cient\'{\i}fico e
Tecnol{\'o}gico and Funda\c c{\~a}o de Amparo {\`a} Pesquisa do
Estado de S{\~a}o Paulo (Brazil),
Natural Science Foundation of China (P.~R.~China),
Ministry of Education, Youth and Sports (Czech Republic),
Centre National de la Recherche Scientifique, Commissariat
{\`a} l'{\'E}nergie Atomique, and Institut National de Physique
Nucl{\'e}aire et de Physique des Particules (France),
Bundesministerium f\"ur Bildung und Forschung, Deutscher
Akademischer Austausch Dienst, and Alexander von Humboldt Stiftung (Germany),
Hungarian National Science Fund, OTKA (Hungary), 
Department of Atomic Energy (India), 
Israel Science Foundation (Israel), 
National Research Foundation and WCU program of the 
Ministry Education Science and Technology (Korea),
Physics Department, Lahore University of Management Sciences (Pakistan),
Ministry of Education and Science, Russian Academy of Sciences,
Federal Agency of Atomic Energy (Russia),
VR and Wallenberg Foundation (Sweden), 
the U.S. Civilian Research and Development Foundation for the
Independent States of the Former Soviet Union, 
the US-Hungarian NSF-OTKA-MTA, 
and the US-Israel Binational Science Foundation.



\begin{thebibliography}{45}
\expandafter\ifx\csname natexlab\endcsname\relax\def\natexlab#1{#1}\fi
\expandafter\ifx\csname bibnamefont\endcsname\relax
  \def\bibnamefont#1{#1}\fi
\expandafter\ifx\csname bibfnamefont\endcsname\relax
  \def\bibfnamefont#1{#1}\fi
\expandafter\ifx\csname citenamefont\endcsname\relax
  \def\citenamefont#1{#1}\fi
\expandafter\ifx\csname url\endcsname\relax
  \def\url#1{\texttt{#1}}\fi
\expandafter\ifx\csname urlprefix\endcsname\relax\def\urlprefix{URL }\fi
\providecommand{\bibinfo}[2]{#2}
\providecommand{\eprint}[2][]{\url{#2}}

\bibitem[{\citenamefont{Karsch}(2002)}]{Karsch:2001cy}
\bibinfo{author}{\bibfnamefont{F.}~\bibnamefont{Karsch}},
  \bibinfo{journal}{Lect. Notes Phys.} \textbf{\bibinfo{volume}{583}},
  \bibinfo{pages}{209} (\bibinfo{year}{2002}).

\bibitem[{\citenamefont{Adare et~al.}(2010)}]{PPG086}
\bibinfo{author}{\bibfnamefont{A.}~\bibnamefont{Adare}} \bibnamefont{{\em et~al.}}
  (\bibinfo{collaboration}{PHENIX Collaboration}), \bibinfo{journal}{Phys. Rev.
  Lett.} \textbf{\bibinfo{volume}{104}}, \bibinfo{pages}{132301}
  (\bibinfo{year}{2010}).

\bibitem[{\citenamefont{Adare et~al.}(2011)}]{PPG077}
\bibinfo{author}{\bibfnamefont{A.}~\bibnamefont{Adare}} \bibnamefont{{\em et~al.}}
  (\bibinfo{collaboration}{PHENIX Collaboration}), \bibinfo{journal}{Phys. Rev.
  C} \textbf{\bibinfo{volume}{84}}, \bibinfo{pages}{044905}
  (\bibinfo{year}{2011}).

\bibitem[{\citenamefont{Adare et~al.}(2007)}]{PPG066}
\bibinfo{author}{\bibfnamefont{A.}~\bibnamefont{Adare}} \bibnamefont{{\em et~al.}}
  (\bibinfo{collaboration}{PHENIX Collaboration}), \bibinfo{journal}{Phys. Rev.
  Lett.} \textbf{\bibinfo{volume}{98}}, \bibinfo{pages}{172301}
  (\bibinfo{year}{2007}).

\bibitem[{\citenamefont{Adare et~al.}(2006)}]{PPG065}
\bibinfo{author}{\bibfnamefont{A.}~\bibnamefont{Adare}} \bibnamefont{{\em et~al.}}
  (\bibinfo{collaboration}{PHENIX Collaboration}), \bibinfo{journal}{Phys. Rev.
  Lett.} \textbf{\bibinfo{volume}{97}}, \bibinfo{pages}{252002}
  (\bibinfo{year}{2006}).

\bibitem[{\citenamefont{Dokshitzer and Kharzeev}(2001)}]{dead_cone}
\bibinfo{author}{\bibfnamefont{Y.~L.} \bibnamefont{Dokshitzer}}
  \bibnamefont{and} \bibinfo{author}{\bibfnamefont{D.}~\bibnamefont{Kharzeev}},
  \bibinfo{journal}{Phys. Lett. B} \textbf{\bibinfo{volume}{519}},
  \bibinfo{pages}{199} (\bibinfo{year}{2001}).

\bibitem[{\citenamefont{Moore and Teaney}(2005)}]{Teaney}
\bibinfo{author}{\bibfnamefont{G.~D.} \bibnamefont{Moore}} \bibnamefont{and}
  \bibinfo{author}{\bibfnamefont{D.}~\bibnamefont{Teaney}},
  \bibinfo{journal}{Phys. Rev. C} \textbf{\bibinfo{volume}{71}},
  \bibinfo{pages}{064904} (\bibinfo{year}{2005}).

\bibitem[{\citenamefont{Gossiaux and Aichelin}(2008)}]{Gossiaux1}
\bibinfo{author}{\bibfnamefont{P.~B.} \bibnamefont{Gossiaux}} \bibnamefont{and}
  \bibinfo{author}{\bibfnamefont{J.}~\bibnamefont{Aichelin}},
  \bibinfo{journal}{Phys. Rev. C} \textbf{\bibinfo{volume}{78}},
  \bibinfo{pages}{014904} (\bibinfo{year}{2008}).

\bibitem[{\citenamefont{van Hees et~al.}(2006)\citenamefont{van Hees, Greco,
  and Rapp}}]{PhysRevC.73.034913}
\bibinfo{author}{\bibfnamefont{H.}~\bibnamefont{van Hees}},
  \bibinfo{author}{\bibfnamefont{V.}~\bibnamefont{Greco}}, \bibnamefont{and}
  \bibinfo{author}{\bibfnamefont{R.}~\bibnamefont{Rapp}},
  \bibinfo{journal}{Phys. Rev. C} \textbf{\bibinfo{volume}{73}},
  \bibinfo{pages}{034913} (\bibinfo{year}{2006}).

\bibitem[{\citenamefont{Gossiaux et~al.}(2009)\citenamefont{Gossiaux,
  Bierkandt, and Aichelin}}]{Gossiaux2}
\bibinfo{author}{\bibfnamefont{P.~B.} \bibnamefont{Gossiaux}},
  \bibinfo{author}{\bibfnamefont{R.}~\bibnamefont{Bierkandt}},
  \bibnamefont{and} \bibinfo{author}{\bibfnamefont{J.}~\bibnamefont{Aichelin}},
  \bibinfo{journal}{Phys. Rev. C} \textbf{\bibinfo{volume}{79}},
  \bibinfo{pages}{044906} (\bibinfo{year}{2009}).

\bibitem[{\citenamefont{Gossiaux and Aichelin}(2009)}]{Gossiaux3}
\bibinfo{author}{\bibfnamefont{P.~B.} \bibnamefont{Gossiaux}} \bibnamefont{and}
  \bibinfo{author}{\bibfnamefont{J.}~\bibnamefont{Aichelin}},
  \bibinfo{journal}{J. Phys. G} \textbf{\bibinfo{volume}{36}},
  \bibinfo{pages}{064028} (\bibinfo{year}{2009}).

\bibitem[{\citenamefont{Mustafa}(2005)}]{PhysRevC.72.014905}
\bibinfo{author}{\bibfnamefont{M.~G.} \bibnamefont{Mustafa}},
  \bibinfo{journal}{Phys. Rev. C} \textbf{\bibinfo{volume}{72}},
  \bibinfo{pages}{014905} (\bibinfo{year}{2005}).

\bibitem[{\citenamefont{Wicks et~al.}(2007{\natexlab{a}})\citenamefont{Wicks,
  Horowitz, Djordjevic, and Gyulassy}}]{wicks_pqcd}
\bibinfo{author}{\bibfnamefont{S.}~\bibnamefont{Wicks}},
  \bibinfo{author}{\bibfnamefont{W.}~\bibnamefont{Horowitz}},
  \bibinfo{author}{\bibfnamefont{M.}~\bibnamefont{Djordjevic}},
  \bibnamefont{and} \bibinfo{author}{\bibfnamefont{M.}~\bibnamefont{Gyulassy}},
  \bibinfo{journal}{Nucl. Phys. A} \textbf{\bibinfo{volume}{783}},
  \bibinfo{pages}{493} (\bibinfo{year}{2007}{\natexlab{a}}).

\bibitem[{\citenamefont{Adil and Vitev}(2007)}]{Adil2007139}
\bibinfo{author}{\bibfnamefont{A.}~\bibnamefont{Adil}} \bibnamefont{and}
  \bibinfo{author}{\bibfnamefont{I.}~\bibnamefont{Vitev}},
  \bibinfo{journal}{Phys. Lett. B} \textbf{\bibinfo{volume}{649}},
  \bibinfo{pages}{139} (\bibinfo{year}{2007}).

\bibitem[{\citenamefont{Sorensen and
  Dong}(2006{\natexlab{a}})}]{PhysRevC.74.024902}
\bibinfo{author}{\bibfnamefont{P.}~\bibnamefont{Sorensen}} \bibnamefont{and}
  \bibinfo{author}{\bibfnamefont{X.}~\bibnamefont{Dong}},
  \bibinfo{journal}{Phys. Rev. C} \textbf{\bibinfo{volume}{74}},
  \bibinfo{pages}{024902} (\bibinfo{year}{2006}{\natexlab{a}}).

\bibitem[{\citenamefont{Eskola et~al.}()\citenamefont{Eskola, Paukkunen, and
  Salgado}}]{EPS09}
\bibinfo{author}{\bibfnamefont{K.~J.} \bibnamefont{Eskola}},
  \bibinfo{author}{\bibfnamefont{H.}~\bibnamefont{Paukkunen}},
  \bibnamefont{and} \bibinfo{author}{\bibfnamefont{C.~A.}
  \bibnamefont{Salgado}}, \bibinfo{note}{J. High Energy Phys. {\bf 04} (2009)
  065}.

\bibitem[{\citenamefont{Vitev}(2007)}]{CNM_Eloss}
\bibinfo{author}{\bibfnamefont{I.}~\bibnamefont{Vitev}},
  \bibinfo{journal}{Phys. Rev. C} \textbf{\bibinfo{volume}{75}},
  \bibinfo{pages}{064906} (\bibinfo{year}{2007}).

\bibitem[{\citenamefont{Adare et~al.}()}]{Adare:2013piz}
\bibinfo{author}{\bibfnamefont{A.}~\bibnamefont{Adare}} \bibnamefont{{\em et~al.}}
  (\bibinfo{collaboration}{PHENIX Collaboration}),
  \bibinfo{note}{arXiv:1303.1794 and to be published}.

\bibitem[{\citenamefont{Chatrchyan et~al.}(2013)}]{Chatrchyan2013795}
\bibinfo{author}{\bibfnamefont{S.}~\bibnamefont{Chatrchyan}}
  \bibnamefont{{\em et~al.}}, \bibinfo{journal}{Phys. Lett. B}
  \textbf{\bibinfo{volume}{718}}, \bibinfo{pages}{795 } (\bibinfo{year}{2013}).

\bibitem[{\citenamefont{Abelev et~al.}(2013)}]{Abelev201329}
\bibinfo{author}{\bibfnamefont{B.}~\bibnamefont{Abelev}} \bibnamefont{{\em et~al.}}
  (\bibinfo{collaboration}{ALICE Collaboration}), \bibinfo{journal}{Phys. Lett.
  B} \textbf{\bibinfo{volume}{719}}, \bibinfo{pages}{29 }
  (\bibinfo{year}{2013}).

\bibitem[{\citenamefont{Aad et~al.}(2013)}]{PhysRevLett.110.182302}
\bibinfo{author}{\bibfnamefont{G.}~\bibnamefont{Aad}} \bibnamefont{{\em et~al.}}
  (\bibinfo{collaboration}{ATLAS Collaboration}), \bibinfo{journal}{Phys. Rev.
  Lett.} \textbf{\bibinfo{volume}{110}}, \bibinfo{pages}{182302}
  (\bibinfo{year}{2013}).

\bibitem[{\citenamefont{Adare et~al.}(2012{\natexlab{a}})}]{PPG131}
\bibinfo{author}{\bibfnamefont{A.}~\bibnamefont{Adare}} \bibnamefont{{\em et~al.}}
  (\bibinfo{collaboration}{PHENIX Collaboration}), \bibinfo{journal}{Phys. Rev.
  Lett.} \textbf{\bibinfo{volume}{109}}, \bibinfo{pages}{242301}
  (\bibinfo{year}{2012}{\natexlab{a}}).

\bibitem[{\citenamefont{Adcox et~al.}(2003)}]{PHENIXNIM}
\bibinfo{author}{\bibfnamefont{K.}~\bibnamefont{Adcox}} \bibnamefont{{\em et~al.}}
  (\bibinfo{collaboration}{PHENIX Collaboration}), \bibinfo{journal}{Nucl.
  Instrum. Methods A} \textbf{\bibinfo{volume}{499}}, \bibinfo{pages}{469}
  (\bibinfo{year}{2003}).

\bibitem[{\citenamefont{Miller et~al.}(2007)\citenamefont{Miller, Reygers,
  Sanders, and Steinberg}}]{glauber}
\bibinfo{author}{\bibfnamefont{M.~L.} \bibnamefont{Miller}},
  \bibinfo{author}{\bibfnamefont{K.}~\bibnamefont{Reygers}},
  \bibinfo{author}{\bibfnamefont{S.~J.} \bibnamefont{Sanders}},
  \bibnamefont{and}
  \bibinfo{author}{\bibfnamefont{P.}~\bibnamefont{Steinberg}},
  \bibinfo{journal}{Ann. Rev. Nucl. Part. Sci.} \textbf{\bibinfo{volume}{57}},
  \bibinfo{pages}{205} (\bibinfo{year}{2007}).

\bibitem[{\citenamefont{Glauber and Matthiae}(1970)}]{Glauber:1970jm}
\bibinfo{author}{\bibfnamefont{R.}~\bibnamefont{Glauber}} \bibnamefont{and}
  \bibinfo{author}{\bibfnamefont{G.}~\bibnamefont{Matthiae}},
  \bibinfo{journal}{Nucl. Phys. B} \textbf{\bibinfo{volume}{21}},
  \bibinfo{pages}{135} (\bibinfo{year}{1970}).

\bibitem[{\citenamefont{Adare et~al.}(2008{\natexlab{a}})}]{PPG084}
\bibinfo{author}{\bibfnamefont{A.}~\bibnamefont{Adare}} \bibnamefont{{\em et~al.}}
  (\bibinfo{collaboration}{PHENIX Collaboration}), \bibinfo{journal}{Phys. Rev.
  Lett.} \textbf{\bibinfo{volume}{101}}, \bibinfo{pages}{162301}
  (\bibinfo{year}{2008}{\natexlab{a}}).

\bibitem[{\citenamefont{Adare et~al.}(2008{\natexlab{b}})}]{PPG071}
\bibinfo{author}{\bibfnamefont{A.}~\bibnamefont{Adare}} \bibnamefont{{\em et~al.}}
  (\bibinfo{collaboration}{PHENIX Collaboration}), \bibinfo{journal}{Phys. Rev.
  Lett.} \textbf{\bibinfo{volume}{101}}, \bibinfo{pages}{122301}
  (\bibinfo{year}{2008}{\natexlab{b}}).

\bibitem[{\citenamefont{Cacciari et~al.}(2005)\citenamefont{Cacciari, Nason,
  and Vogt}}]{FONLL}
\bibinfo{author}{\bibfnamefont{M.}~\bibnamefont{Cacciari}},
  \bibinfo{author}{\bibfnamefont{P.}~\bibnamefont{Nason}}, \bibnamefont{and}
  \bibinfo{author}{\bibfnamefont{R.}~\bibnamefont{Vogt}},
  \bibinfo{journal}{Phys. Rev. Lett.} \textbf{\bibinfo{volume}{95}},
  \bibinfo{pages}{122001} (\bibinfo{year}{2005}).

\bibitem[{\citenamefont{Adler et~al.}(2007)}]{PPG044}
\bibinfo{author}{\bibfnamefont{S.~S.} \bibnamefont{Adler}} \bibnamefont{{\em et~al.}}
  (\bibinfo{collaboration}{PHENIX Collaboration}), \bibinfo{journal}{Phys. Rev.
  Lett.} \textbf{\bibinfo{volume}{98}}, \bibinfo{pages}{172302}
  (\bibinfo{year}{2007}).

\bibitem[{\citenamefont{Adare et~al.}(2013)}]{PPG146}
\bibinfo{author}{\bibfnamefont{A.}~\bibnamefont{Adare}} \bibnamefont{{\em et~al.}}
  (\bibinfo{collaboration}{PHENIX Collaboration}), \bibinfo{journal}{Phys. Rev.
  C} \textbf{\bibinfo{volume}{88}}, \bibinfo{pages}{024906}
  (\bibinfo{year}{2013}).

\bibitem[{\citenamefont{Cronin et~al.}(1975)\citenamefont{Cronin, Frisch,
  Shochet, Boymond, Pirou\'e, and Sumner}}]{Cronin}
\bibinfo{author}{\bibfnamefont{J.~W.} \bibnamefont{Cronin}},
  \bibinfo{author}{\bibfnamefont{H.~J.} \bibnamefont{Frisch}},
  \bibinfo{author}{\bibfnamefont{M.~J.} \bibnamefont{Shochet}},
  \bibinfo{author}{\bibfnamefont{J.~P.} \bibnamefont{Boymond}},
  \bibinfo{author}{\bibfnamefont{P.~A.} \bibnamefont{Pirou\'e}},
  \bibnamefont{and} \bibinfo{author}{\bibfnamefont{R.~L.}
  \bibnamefont{Sumner}}, \bibinfo{journal}{Phys. Rev. D}
  \textbf{\bibinfo{volume}{11}}, \bibinfo{pages}{3105} (\bibinfo{year}{1975}).

\bibitem[{\citenamefont{Accardi}()}]{Accardi}
\bibinfo{author}{\bibfnamefont{A.}~\bibnamefont{Accardi}},
  \bibinfo{note}{arXiv:hep-ph/0212148}.

\bibitem[{\citenamefont{Hwa and Yang}(2004{\natexlab{a}})}]{Hwa1}
\bibinfo{author}{\bibfnamefont{R.~C.} \bibnamefont{Hwa}} \bibnamefont{and}
  \bibinfo{author}{\bibfnamefont{C.~B.} \bibnamefont{Yang}},
  \bibinfo{journal}{Phys. Rev. C} \textbf{\bibinfo{volume}{70}},
  \bibinfo{pages}{024905} (\bibinfo{year}{2004}{\natexlab{a}}).

\bibitem[{\citenamefont{Hwa and Yang}(2004{\natexlab{b}})}]{Hwa2}
\bibinfo{author}{\bibfnamefont{R.~C.} \bibnamefont{Hwa}} \bibnamefont{and}
  \bibinfo{author}{\bibfnamefont{C.~B.} \bibnamefont{Yang}},
  \bibinfo{journal}{Phys. Rev. Lett.} \textbf{\bibinfo{volume}{93}},
  \bibinfo{pages}{082302} (\bibinfo{year}{2004}{\natexlab{b}}).

\bibitem[{\citenamefont{Sorensen and
  Dong}(2006{\natexlab{b}})}]{charm_chemistry}
\bibinfo{author}{\bibfnamefont{P.}~\bibnamefont{Sorensen}} \bibnamefont{and}
  \bibinfo{author}{\bibfnamefont{X.}~\bibnamefont{Dong}},
  \bibinfo{journal}{Phys. Rev. C} \textbf{\bibinfo{volume}{74}},
  \bibinfo{pages}{024902} (\bibinfo{year}{2006}{\natexlab{b}}).

\bibitem[{\citenamefont{Helenius et~al.}()\citenamefont{Helenius, Eskola,
  Honkanen, and Salgado}}]{Helenius}
\bibinfo{author}{\bibfnamefont{I.}~\bibnamefont{Helenius}},
  \bibinfo{author}{\bibfnamefont{K.~J.} \bibnamefont{Eskola}},
  \bibinfo{author}{\bibfnamefont{H.}~\bibnamefont{Honkanen}}, \bibnamefont{and}
  \bibinfo{author}{\bibfnamefont{C.~A.} \bibnamefont{Salgado}},
  \bibinfo{note}{arXiv:1205.5359}.

\bibitem[{\citenamefont{Sharma et~al.}(2009)\citenamefont{Sharma, Vitev, and
  Zhang}}]{SharmaVitev}
\bibinfo{author}{\bibfnamefont{R.}~\bibnamefont{Sharma}},
  \bibinfo{author}{\bibfnamefont{I.}~\bibnamefont{Vitev}}, \bibnamefont{and}
  \bibinfo{author}{\bibfnamefont{B.-W.} \bibnamefont{Zhang}},
  \bibinfo{journal}{Phys. Rev. C} \textbf{\bibinfo{volume}{80}},
  \bibinfo{pages}{054902} (\bibinfo{year}{2009}).

\bibitem[{\citenamefont{Wicks et~al.}(2007{\natexlab{b}})\citenamefont{Wicks,
  Horowitz, Djordjevic, and Gyulassy}}]{Wicks2007426}
\bibinfo{author}{\bibfnamefont{S.}~\bibnamefont{Wicks}},
  \bibinfo{author}{\bibfnamefont{W.}~\bibnamefont{Horowitz}},
  \bibinfo{author}{\bibfnamefont{M.}~\bibnamefont{Djordjevic}},
  \bibnamefont{and} \bibinfo{author}{\bibfnamefont{M.}~\bibnamefont{Gyulassy}},
  \bibinfo{journal}{Nucl. Phys. A} \textbf{\bibinfo{volume}{784}},
  \bibinfo{pages}{426 } (\bibinfo{year}{2007}{\natexlab{b}}).

\bibitem[{\citenamefont{Zhang et~al.}(2004)\citenamefont{Zhang, Wang, and
  Wang}}]{zhang_pqcd}
\bibinfo{author}{\bibfnamefont{B.~W.}~\bibnamefont{Zhang}},
  \bibinfo{author}{\bibfnamefont{E.}~\bibnamefont{Wang}}, \bibnamefont{and}
  \bibinfo{author}{\bibfnamefont{X.~N.}~\bibnamefont{Wang}},
  \bibinfo{journal}{Phys. Rev. Lett.} \textbf{\bibinfo{volume}{93}},
  \bibinfo{pages}{072301} (\bibinfo{year}{2004}).

\bibitem[{\citenamefont{Dominguez and Wu}(2009)}]{Dominguez2009246}
\bibinfo{author}{\bibfnamefont{F.}~\bibnamefont{Dominguez}} \bibnamefont{and}
  \bibinfo{author}{\bibfnamefont{B.}~\bibnamefont{Wu}}, \bibinfo{journal}{Nucl.
  Phys. A} \textbf{\bibinfo{volume}{818}}, \bibinfo{pages}{246}
  (\bibinfo{year}{2009}).

\bibitem[{\citenamefont{Dominguez et~al.}(2009)\citenamefont{Dominguez,
  Marquet, and Wu}}]{Dominguez200999}
\bibinfo{author}{\bibfnamefont{F.}~\bibnamefont{Dominguez}},
  \bibinfo{author}{\bibfnamefont{C.}~\bibnamefont{Marquet}}, \bibnamefont{and}
  \bibinfo{author}{\bibfnamefont{B.}~\bibnamefont{Wu}}, \bibinfo{journal}{Nucl.
  Phys. A} \textbf{\bibinfo{volume}{823}}, \bibinfo{pages}{99 }
  (\bibinfo{year}{2009}).

\bibitem[{\citenamefont{Wong}(2005)}]{PhysRevC.72.034906}
\bibinfo{author}{\bibfnamefont{C.-Y.} \bibnamefont{Wong}},
  \bibinfo{journal}{Phys. Rev. C} \textbf{\bibinfo{volume}{72}},
  \bibinfo{pages}{034906} (\bibinfo{year}{2005}).

\bibitem[{\citenamefont{Adare et~al.}(2012{\natexlab{b}})}]{PPG117}
\bibinfo{author}{\bibfnamefont{A.}~\bibnamefont{Adare}} \bibnamefont{{\em et~al.}}
  (\bibinfo{collaboration}{PHENIX Collaboration}), \bibinfo{journal}{Phys. Rev.
  C} \textbf{\bibinfo{volume}{86}}, \bibinfo{pages}{024909}
  (\bibinfo{year}{2012}{\natexlab{b}}).

\bibitem[{\citenamefont{Bjorken}(1983)}]{Bjorken}
\bibinfo{author}{\bibfnamefont{J.~D.} \bibnamefont{Bjorken}},
  \bibinfo{journal}{Phys. Rev. D} \textbf{\bibinfo{volume}{27}},
  \bibinfo{pages}{140} (\bibinfo{year}{1983}).

\bibitem[{\citenamefont{Abelev et~al.}(2012)}]{ALICE}
\bibinfo{author}{\bibfnamefont{B.}~\bibnamefont{Abelev}} \bibnamefont{{\em et~al.}}
  (\bibinfo{collaboration}{ALICE Collaboration}), \bibinfo{journal}{Phys. Rev.
  D} \textbf{\bibinfo{volume}{86}}, \bibinfo{pages}{112007}
  (\bibinfo{year}{2012}).

\end{thebibliography}

\end{document}